\documentclass[a4paper, amsfonts, amssymb, amsmath, reprint, showkeys, nofootinbib, twoside, superscriptaddress]{revtex4-1}
\usepackage[english]{babel}
\usepackage[utf8]{inputenc}
\usepackage[colorinlistoftodos, color=green!40, prependcaption]{todonotes}
\usepackage{amsthm}
\usepackage{mathtools}
\usepackage{physics}
\usepackage{xcolor}
\usepackage{graphicx}
\usepackage[left=23mm,right=13mm,top=35mm,columnsep=15pt]{geometry} 
\usepackage{adjustbox}
\usepackage{placeins}
\usepackage[T1]{fontenc}
\usepackage{lipsum}
\usepackage{csquotes}
\usepackage{siunitx}
\usepackage{caption}
\usepackage{subcaption}
\usepackage[pdftex, pdftitle={Article}, pdfauthor={Author}, colorlinks=true]{hyperref}
\begin{document}
\title{ 
Micron-size spatial superpositions for the QGEM-protocol via screening and trapping
}

\author{Martine Schut}
    \affiliation{Van Swinderen Institute for Particle Physics and Gravity, University of Groningen, 9747AG Groningen, the Netherlands }
    \affiliation{Bernoulli Institute for Mathematics, Computer Science and Artificial Intelligence, University of Groningen, 9747 AG Groningen, the Netherlands \vspace{1mm}}

       \author{Andrew Geraci}
    \affiliation{Department of Physics and Astronomy, Northwestern University, 2145 Sheridan Road, Evanston, IL}
  \author{Sougato Bose}
    \affiliation{Department of Physics and Astronomy, University College London, London WC1E 6BT, United Kingdom}
\author{Anupam Mazumdar }
    \affiliation{Van Swinderen Institute for Particle Physics and Gravity, University of Groningen, 9747AG Groningen, the Netherlands }


\begin{abstract}
The quantum gravity-induced entanglement of masses (QGEM) protocol for testing quantum gravity using entanglement witnessing utilizes the creation of spatial quantum superpositions of two neutral, massive matter-wave interferometers kept adjacent to each other, separated by a distance $d$. 
The mass and the spatial superposition should be such that the two quantum systems can entangle solely via the quantum nature of gravity. 
Despite being charge-neutral, there are many electromagnetic backgrounds that can also entangle the systems, such as the dipole-dipole interaction, and the Casimir-Polder interaction. 
To minimize electromagnetic-induced interactions between the masses it is pertinent to isolate the two superpositions by a conducting plate. 
However, the conducting plate will also exert forces on the masses and hence the trajectories of the two superpositions would be affected. 
To minimize this effect, we propose to trap the two interferometers such that the trapping potential dominates over the attraction between the conducting plate and the matter-wave interferometers. 
The superpositions can still be created via the Stern-Gerlach effect in the direction parallel to the plate, where the trapping potential is negligible.
The combination of trapping and shielding provides a better parameter space for the parallel configuration of the experiment, where the requirement on the size of the spatial superposition, to witness the entanglement between the two masses purely due to their quantum nature of gravity, decreases by at least two orders of magnitude as compared to the original protocol paper~\cite{Bose:2017nin}.
\end{abstract}


\maketitle

\section{Introduction}\label{sec:intro}
One of the biggest challenges in theoretical physics is to test the nature of gravity, whether gravity obeys the rules of quantum mechanics or not. As we know, it is extremely hard to detect the carrier of the gravitational interaction, known as the graviton, a massless spin-2 quanta~\cite{dyson_is_2013}. However, a protocol known as the quantum-gravity induced entanglement of masses (QGEM)~\cite{Bose:2017nin}~\footnote{See also~\cite{Marletto:2017kzi}, which appeared on the same day as~\cite{Bose:2017nin}.
The original results  of \cite{Bose:2017nin} were first reported in the conference talk in Bangalore \cite{ICTS}
}, proposes a table-top experiment that exploits fundamentally quantum properties such as quantum superposition and quantum entanglement to witness the quantum nature of gravity. 
The quantum nature of gravity can be evidenced by witnessing the quantum entanglement at the level of Newtonian potential despite $\hbar$ factors canceling out in the observable. This is due to the fact that even at the Newtonian level the gravitational interaction is being mediated by the virtual excitations of the massless spin-2 graviton, see~\cite{Bose:2017nin,Marshman:2019sne,Bose:2022uxe}, and ~\cite{Danielson:2021egj,Carney_2019,Carney:2021vvt,Christodoulou:2022vte}. 
Due to the quantum  nature of the gravitational interaction, the potential is an operator-valued entity, as shown in~\cite{Bose:2022uxe,Vinckers:2023grv}, see also textbook~\cite{Scadron:1991ep}, and noted reviews on this topics~\cite{Donoghue:1994dn,Donoghue:2012zc}.

The QGEM's observation is very similar to Bell's original idea of testing quantum mechanics~\cite{Hensen:2015ccp}, and also the observation made in Refs.~\cite{PhysRevA.46.4413,GISIN199215} that the quantum correlation exists for the large angular momentum systems despite $\hbar \rightarrow 0$ limit. Naturally, the detected entanglement can only be generated by quantum interactions between the test particles, according to the Local Operations and Classical Communication (LOCC) principle~\cite{Bennett:1996gf}. Hence, only if gravity is a quantum entity will it generate the entanglement between the two spatially quantum superposed masses~\cite{Bose:2017nin,Bose:2022uxe,Vinckers:2023grv}. 

A similar protocol involving photon and matter entanglement due to the quantum nature of gravity reveals the quantum properties of the spin-2 nature of the graviton~\cite{Biswas:2022qto}. 
Furthermore, a QGEM can test the quantum version of the equivalence principle~\cite{Bose:2022czr}, potentially test the fundamental non-locality in quantum gravity~\cite{Vinckers:2023grv}, and test theories that possess a massive graviton~\cite{Elahi:2023ozf}. Recently, a complementary probe to the quantum nature of gravity has been proposed in the context of the measurement process of the quantum state of matter in the presence of a quantum gravitational interaction~\cite{Hanif:2023fto}.

Given the import of this protocol it is pertinent to study the realization of the experimental protocol discussed in \cite{Bose:2017nin}. 
The QGEM protocol relies on the creation of two massive spatial superpositions kept adjacent to each other.
Introducing a minimum distance between the two neutral masses should take into account any electromagnetic interactions, such as the Casimir-Polder interaction, and dipole-dipole interactions, as they cause a major source of decoherence for the matter-wave interferometers, see the discussions in~\cite{vandeKamp:2020rqh,Schut:2023eux,Fragolino:2023agd}. 
It was suggested in~\cite{Bose:2017nin} that the two masses of $\sim 10^{-14}\,\si{\kilogram}$, in spatial superposition of width $\Delta x \sim 250\,\si{\micro\metre}$ and separated by a minimal distance $d\sim 200\,\si{\micro\metre}$ (see fig.~\ref{fig:setup_lin}) in a time $\sim 2 \,\si{\s}$ will generate a detectable quantum gravity-induced entanglement phase.

The creation of a spatial superposition  can be achieved by considering the test masses to be diamond-like microspheres with NV-centre defects~\cite{Bose:2017nin}, where a spin is embedded in the NV center of each test mass, which can interact with an external inhomogeneous magnetic field. 
Using the coupling of the inhomogeneous magnetic field to the spin, one can create a spatial superposition, similar to in the Stern-Gerlach (SG) apparatus, see ~\cite{Bose:2017nin,  PhysRevLett.123.083601,Margalit:2020qcy,Marshman:2021wyk,PhysRevLett.125.023602,Zhou:2022frl,Zhou:2022jug,Zhou:2022epb,Marshman:2023nkh,PhysRevA.105.012824}.
Similarly, the SG apparatus can be used to reverse the spatial splitting to close the one-loop interferometer. 
Spin measurements of repeated runs are necessary to build the entanglement witness~\cite{Horodecki:2009zz}, which can determine the generation of the entanglement with a certainty dependent on the witness (generally taken to be the Partial Positive Transpose witness, see~\cite{Chevalier:2020uvv,Tilly:2021qef,Schut:2021svd}) and number of measurements.

Many outstanding challenges arise from the entanglement phase being very small compared to the noise which induces dephasing/decoherence. 
Besides the above-mentioned noise such as the gravity-gradient and relative acceleration noise~\cite{Grossardt:2020def,Toros:2020krn,Wu:2022rdv}, there is a dephasing due to heavy massive object near the experiment, e.g., cryogenic devices and vacuum pump~\cite{Gunnink:2022ner}. There is also decoherence due to the heating of the crystal and the interaction with the black body photons and left-over air molecules in the vacuum chamber~\cite{Bose:2017nin,vandeKamp:2020rqh,Nguyen:2019huk,Schut:2021svd,Tilly:2021qef,Rijavec:2020qxd}. 
The decoherence rates are based on earlier computations of~\cite{Romero_Isart_2011,Chang_2009,Sinha:2022snc}.
Next to that, although the test masses are considered neutral objects, there are also electromagnetic interactions between the test masses and between the test mass and its environment~\cite{vandeKamp:2020rqh,Marshman:2023nkh,Fragolino:2023agd,Schut:2023eux}, such as Casimir-Polder and dipole-dipole interactions. 
These electromagnetic interactions also decohere the matter-wave interferometer. Furthermore we have to ensure that we obtain a pure state to initalise the experiment. There are experimental efforts to obtain the ground state of a quantum system, in particular the motional ground state~\cite{Deli__2020,Tebbenjohanns_2020}, and exceptionally heavy object such as one obtained by the LIGO collaboration~\cite{Whittle:2021mtt}. These examples may be considered as masses can be prepared close to the a pure state.

However, above all these experimental constraints, the most daunting task is yet to create a macroscopic spatial quantum superposition. 
In this paper, we address the requirement on spatial superposition by introducing the idea of electromagnetic screening in combination with the trapping of the particle to optimize the entanglement phase due to the quantum nature of the graviton. 
The introduction of the electromagnetic screening in the QGEM setup was first introduced in~\cite{vandeKamp:2020rqh}, and then in~\cite{Schut:2023eux} where the authors studied various dephasing/decoherence effects.

It was noticed earlier that an alternative configuration called the `parallel' configuration of the test masses (fig.~\ref{fig:setup_par}) improves the entanglement generation and improves the entanglement witness~\cite{Nguyen:2019huk}, see~\cite{Schut:2021svd,Tilly:2021qef,Rijavec:2020qxd}.
The effect of screening in a parallel setup was therefore explored to investigate the common mode rejection~\cite{Schut:2023eux}, where the dipole-dipole and Casimir-Polder interaction between the test masses and the plate were considered. 
The conducting plate that separated the two test masses is also responsible for exerting forces on the two halves of the quantum superposition~\cite{vandeKamp:2020rqh,Schut:2021svd}, resulting in a setup where the trajectories of the superposition were drifted towards the conducting plate. 
To minimize the effect of attraction towards the conducting plate, we will consider a new setup here where we will consider creating the superposition in a magnetic trapped potential, see~\cite{Marshman:2023nkh}, and we optimize the size of the superpositions. 
We will find that the combination of shielding and trapping improves the entanglement phase, and relaxes the experimental constraints that were imposed to witness the entanglement in the original setup by almost 2 orders of magnitude~\cite{Bose:2017nin}.

\section{Linear vs parallel experimental setup}\label{sec:setup}
We consider two test masses with an embedded electronic spin in each (often taken to be the NV-centered diamond crystal). To create the superposition, we take a simple route and divide the trajectories into three phases, similar to~\cite{Bose:2017nin}~\footnote{There are other schemes as well to use the optomechanical setups such as \cite{Miao:2019pxw,Biswas:2022qto} to test the quantum nature of gravity in a lab. In this paper we are solely interested in studying the non-Gaussian Schr\"odinger Cat states with the spatial superpositions.}: 
\textit{Phase 1}: The creation of a spatial superposition from a spin superposition. We assume that the spin is aligned with the NV axis and that is in the $\hat{x}$-direction. 
The embedded spin in the test masses can be brought into a superposition by applying a $\pi/2$ pulse. 
Using the coupling between spin and magnetic field, by applying a magnetic field for a time $\tau_a$ the spin superposition becomes a spatial superposition in the presence of the inhomogeneous magnetic field of the SG apparatus. 
The width of the spatial superposition, $\Delta x$, depends on the type of material, the magnetic field applied, and the time $\tau_a$.
\textit{Phase 2}: After the magnetic field that creates the spatial superposition has been applied, the two test masses interact for a time $\tau$. 
\textit{Phase 3}: Again a magnetic field is applied for a time $\tau_a$, this time to recombine the spatial superposition into the spin superposition. 
Variations of this scheme have been constructed to create superposition via different schemes~\cite{Marshman:2021wyk,Zhou:2022frl,Marshman:2023nkh} to get a very large superposition size of order $1000\,{\rm \mu m}$, 
see~\cite{Zhou:2022jug,Zhou:2022epb}.

During the total time $\tau + 2\tau_a$, the initially non-entangled systems become entangled if the gravitational interaction is quantum.
The initially separable system:
\begin{equation}\label{WF-0}
    \Psi(t=0) = \frac{1}{2} \left( \ket{\uparrow} + \ket{\downarrow} \right)_1 \otimes \left( \ket{\uparrow} + \ket{\downarrow} \right)_2 \, ,
\end{equation}
becomes a non-separable system:
\begin{align}\label{WF-1}
    \Psi(\tau) = \frac{e^{i\phi}}{2} \Big(& \ket{\uparrow}_1\ket{\uparrow}_2 + e^{i\phi_1} \ket{\downarrow}_1\ket{\uparrow}_2 \nonumber \\ &+ e^{i\phi_2} \ket{\uparrow}_1\ket{\downarrow}_2 + \ket{\downarrow}_1 \ket{\uparrow}_2 \Big) \, ,
\end{align}
where the phase is picked up via the quantum gravitational interaction.
The phase of each superposition instance is given by $\phi \sim S/\hbar = U \tau /\hbar$ with $U$ the gravitational potential.
In the gravitational potential, the largest contribution comes from the tree-level exchange of a virtual graviton between the test masses, in the non-relativistic regime of a perturbative quantum gravity theory in the weak field limit, which gives the operator-valued potential 
$\hat{U} = G m^2/\hat{r}$, as argued in ~\cite{Bose:2017nin,Marshman:2019sne,Bose:2022uxe}.
As shown in~\cite{Bose:2022uxe}, only a quantum gravity interaction results in the generation of entanglement.
Since the potential is dependent on the distance, the phase and thus the entanglement are dependent on the configuration of the test masses. 
Figures~\ref{fig:setup_par} and~\ref{fig:setup_lin} show the two possible configurations named parallel (denoted here by `par') and linear (denoted here by `lin'), respectively.
The phases $\phi_1, \phi_2$ picked up by the parallel and linear configurations, respectively, are:
\begin{align}
    \phi_1^\text{par} &= \phi_2^\text{par} = \frac{Gm^2}{\sqrt{d^2+(\Delta x)^2}} \frac{\tau}{\hbar} - \phi^\text{par} \, , \label{eq:phases} \\
    \phi_1^\text{lin} &= \frac{Gm^2}{d} \frac{\tau}{\hbar} - \phi^\text{lin} \, , \qq{} \phi_2^\text{lin} = \frac{Gm^2}{d+2\Delta x} \frac{\tau}{\hbar} - \phi^\text{lin} \, , \label{eq:phases_lin}
\end{align}
with the global spin $\phi$ given by:
\begin{align}
    \phi^\text{par} = \frac{Gm^2}{d} \frac{\tau}{\hbar} \, , \qq{}
    \phi^\text{lin} = \frac{Gm^2}{d+\Delta x} \frac{\tau}{\hbar} \, .
\end{align}
\begin{figure}[t]
\centering
\begin{tikzpicture}
\draw[black, dashed, thick] (0,-1) -- (0,-0.1);
\draw[black, dashed, thick] (1,-1) -- (1,-0.1);
\draw[black, thick] (0,-1.7) -- (0,-1.9) |- (0,-1.8) -- (1,-1.8) node[pos=0.5, below] {$d$} -| (1,-1.7) -- (1,-1.9);
\draw[black, thick] (-0.6,-1) -- (-0.4,-1) |- (-0.5,-1) -- (-0.5,0) node[pos=0.5, left] {$\Delta x$} -| (-0.6,0) -- (-0.4,0);
\filldraw[black]
(0,-1) circle (2pt) node[align=center, below] {\hspace{2mm}$\ket{\uparrow}_1$}
(1,-1) circle (2pt) node[align=center, below] {\hspace{2mm}$\ket{\uparrow}_2$};
\draw[black]
(0,0) circle (2pt) node[align=center, above] {\hspace{2mm}$\ket{\downarrow}_1$}
(1,0) circle (2pt) node[align=center, above] {\hspace{2mm}$\ket{\downarrow}_2$};
\end{tikzpicture}
    \caption{Two test masses of mass $m$, labelled $1$ and $2$, in the parallel configuration. The superposition width is $\Delta x$, and the distance between the $\ket{\uparrow}$-states is $d$.}
    \label{fig:setup_par}
\end{figure}
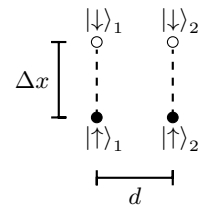
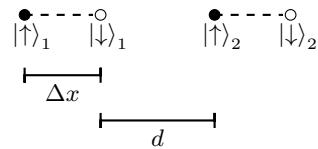
\begin{figure}[t]
\centering
\begin{tikzpicture}
\draw[black, dashed, thick] (-3,0) -- (-2.1,0);
\draw[black, dashed, thick] (-0.5,0) -- (0.4,0);
\draw[black, thick] (-2,-1.3) -- (-2,-1.5) |- (-2,-1.4) -- (-0.5,-1.4) node[pos=0.5, below] {$d$} -| (-0.5,-1.3) -- (-0.5,-1.5);
\draw[black, thick] (-3,-0.7) -- (-3,-0.9) |- (-3,-0.8) -- (-2,-0.8) node[pos=0.5, below] {$\Delta x$} -| (-2,-0.7) -- (-2,-0.9);
\filldraw[black] 
(-3,0) circle (2pt) node[align=center, below] {\hspace{2mm}$\ket{\uparrow}_1$}
(-0.5,0) circle (2pt) node[align=center, below] {\hspace{2mm}$\ket{\uparrow}_2$};
\draw[black]
(-2,0) circle (2pt) node[align=center, below] {\hspace{2mm}$\ket{\downarrow}_1$}
(0.5,0) circle (2pt) node[align=center, below] {\hspace{2mm}$\ket{\downarrow}_2$};
\end{tikzpicture}
\caption{Two test masses of mass $m$, labelled $1$ and $2$, in the linear configuration. The superposition width is $\Delta x$, and the distance between the $\ket{\uparrow}$-states is $d + \Delta x$.}
\label{fig:setup_lin}
\end{figure}
More precisely, the total accumulated entanglement phase is given by $$ \int_0^{\tau+2\tau_a} \dd{t} (\omega_1(t)+\omega_2(t)),$$ where $\omega_{1,2}\,\tau \equiv \phi_{1,2}$ and the time-dependence comes from the time-dependence in creating $\Delta x$. 
However, as an approximation we consider $\Delta x$ here to be time-independent, i.e. only the entanglement generated during the \textit{phase 2} during which the superposition width can be considered constant over a time $\tau$, is taken.

To quantify the entanglement we consider the negativity, for which we need to find the eigenvalues of the partial transpose of the density matrix.
These eigenvalues, for a density matrix $\rho = \ket{\Psi(t)}\bra{\Psi(t)}$ found from eq.~(\ref{WF-1}), are given by:
\begin{align}
    \lambda_{1,2} &= \pm \frac{1}{2} \sin(\frac{\phi_1 +\phi_2}{2}) \label{eq:eigval1}\\
    \lambda_{3,4} &= \frac{1}{2} \pm \frac{1}{2} \cos(\frac{\phi_1 +\phi_2}{2}) \label{eq:eigval2} \, ,
\end{align}
where $\phi_{1,2}$ are dependent on the configuration and given in eqs.~\eqref{eq:phases},~\eqref{eq:phases_lin}.
The eigenvalues $\lambda_{3,4} \geq 0$ always, while $\lambda_1 = - \lambda_2$ indicates that always: $\lambda_1$ or $\lambda_2 \leq 0$.
Thus the negativity, ${\cal N}$, defined as the absolute value of the sum of all negative eigenvalues of the partial transpose of the density matrix, can be expressed as:
\begin{equation}\label{eq:negativity}
    \mathcal{N} = \frac{1}{2} \abs{\sin(\frac{\phi_1 +\phi_2}{2})} \, .
\end{equation}
For $d>0$ and $\Delta x > 0$, the $\phi_1+\phi_2 >0$ is always positive for the linear setup, while it is always negative for the parallel setup~\footnote{
This is obvious from the definition of the phases for the parallel setup. For the linear case one can simply solve the inequality $\phi_1^\text{lin}+\phi_2^\text{lin}>0$ and find that it holds for $\Delta x, d > 0$.}.
This means that due to the absolute signs in eq.~\eqref{eq:negativity}, this expression is applicable to both setups.
In general, it is found that the parallel setup~\cite{Nguyen:2019huk} outperforms the linear setup in terms of entanglement generation for the timescale considered in the QGEM proposal~\cite{Tilly:2021qef,Schut:2021svd,Schut:2023eux,Miki:2020hvg}.
From eq.~\eqref{eq:negativity} we can also see that since $\phi_1^\text{lin}+\phi_2^\text{lin}<\phi_1^\text{par}+\phi_2^\text{par}$ (a solvable inequality that holds for any $\Delta x, d >0$), at small times the parallel setup generates more entanglement than the linear setup.

In the original proposal~\cite{Bose:2017nin}, the linear setup was considered, and follow-up papers often considered this setup as well.
For example, reducing the required separation by introducing a conducting screen was considered in the linear setup~\cite{vandeKamp:2020rqh}.
Here, we will consider the same idea of screening the electromagnetic interactions and additionally we will consider the trapping of the test masses in the presence of a conducting screen. 
However, we will perform our analysis in a parallel setup.
In particular, we will show that the experimental parameters required to measure an entanglement phase of $\order{10^{-1}}-\order{10^{-2}}$ are greatly relaxed when considering the trapped parallel setup with electromagnetic screening. These numbers have been set by the control on the decoherence/dephasing arising from the gravity gradient noise and relative acceleration  noise and other sources of decoherence.

\section{Shielding \& trapping}\label{sec:screening_trapping}
Considering now the parallel superposition, which optimizes the gravitational entanglement phase for the timescale relevant to the QGEM experiment, there are ways to further improve the rate of entanglement generation.

\begin{figure}[b]
    \centering
    \includegraphics[width=\linewidth]{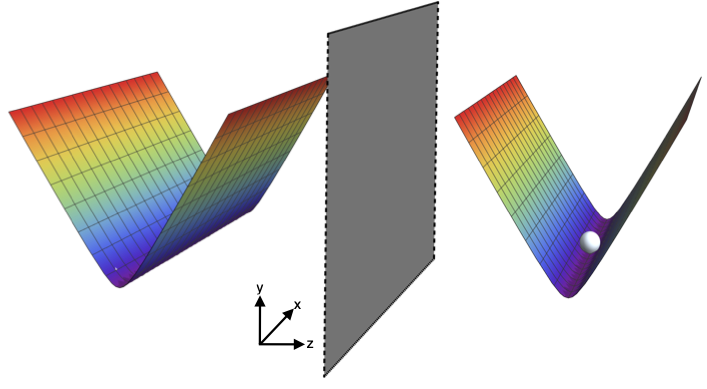}
    \caption{We show a schematic drawing of the proposed setup which involves trapping the test masses and using electromagnetic screening.
    The grey plane represents the conducting plate and it is assumed to be clamped. The rainbow-colored shapes represent the trapping potentials, a spherical test mass can be seen in the right trap. 
    Note that on top of this trapping potential, there will be the magnetic field contribution from the SG apparatus by impinging the transient magnetic field to create the superposition based on \cite{Marshman:2023nkh}, which will produce a weak confinement in the $\hat x$-direction.  
    The trapping potential can also create a small departure from the flat potential along the $\hat x$-direction, see~\cite{Hsu:2016}.
    As an example in this figure, we have plotted the trapping potential of Ref.~\cite{Hsu:2016} for $x\in[-10,10]\,\si{\micro\metre}$, $z=\in[-1,1]\,\si{\micro\metre}$ at $y=0$ (see eq.~\eqref{eq:trapping_profile}).}
    \label{fig:scheme_setup}
\end{figure}

\subsection{Electromagnetic shielding}\label{subsec:shielding}

In general, the test masses used in the QGEM experiment are considered to be diamond microspheres, because they can contain an embedded spin in their NV-centre which has nice properties regarding the spin coherence~\cite{WoodPRB22_GM} and optically read-out properties~\cite{Doherty_2013}. 

A micron size diamond may also have  a permanent dipole of the order $10^{-2}e\,\si{\centi\metre}$ (with $e$ the fundamental electric charge unit) for spheres of size $\sim 10\,\si{\micro\metre}$~\cite{Afek:2021bua,Rider:2016xaq}. Therefore, the two diamond spheres can interact via the dipole-dipole interaction~\cite{Barker:2022mdz,Marshman:2023nkh}. 
In the presence of an electric field, it could also have an induced dipole moment, although this is generally thought to be relatively small for these types of test masses with respect to the permanent dipole, see the analysis of Ref.~\cite{Schut:2023eux}.
Since diamond is a dielectric material, it also interacts via the Casimir-Polder-interaction~\cite{Casimir:1947kzi,Casimir:1948dh}. Hence, entanglement between the two test masses can thus be generated by the dipole-dipole (DD) and by the Casimir-Polder (CP) interaction, which is mediated by the virtual photon.
The interaction potentials between the two spheres (denoted S-S) are given by~\cite{Casimir:1947kzi,griffiths2005introduction,feynman1963feynman}:~\footnote{
In the expression of the Casimir-Polder potential, we have assumed that the separation between the test masses is large compared to the radius of the spheres.
Furthermore, the test masses are assumed to be perfectly spherically and to consist solely of diamonds.
In a high vacuum environment, the polarizability of a diamond sphere can be expressed as $\alpha=R^3(\varepsilon-1)/(\varepsilon+2)$~\cite{kim2005static}. 
\label{footnote:cp}}
\begin{align}
    V_\text{DD}^\text{S-S} &= \frac{1}{4\pi\varepsilon_0 } \left( \frac{\boldsymbol{d}_1\boldsymbol{d}_2}{r^3} - \frac{3(\boldsymbol{d}_1\cdot\boldsymbol{r})(\boldsymbol{d}_2\cdot\boldsymbol{r})}{r^5} \right) \, , \label{eq:dipole_SS}\\
    V_\text{CP}^\text{S-S} &= - \frac{23\hbar c}{4\pi} \left(\frac{\varepsilon-1}{\varepsilon+2} \right)^2 \frac{R^6}{r^7} \, , \label{eq:casimir_SS}
\end{align}
where $\varepsilon$ the dielectric constant of the test mass,  $\varepsilon_0$ is the permeability of space, $\hbar$ is the reduced Planck constant, and $c$ the speed of light. Furthermore, $r$ is the separation between the two spheres, $R$ is the radius of the test mass (which we assume to be perfect spheres), and $\boldsymbol{d}_1$ ($\boldsymbol{d}_2$) the electric dipole moment of test mass $1$ ($2$) which can consist of both a permanent electric dipole moment and an induced electric dipole moment in the case of diamond test masses. 
See also~\cite{Schut:2023eux,Marshman:2023nkh,PhysRevLett.125.023602,vandeKamp:2020rqh} for discussions of these interactions in the context of the QGEM experiment.

Typically, a minimal distance between the test masses is introduced such that the gravitational-induced interaction dominates by some predetermined factor over the dipole-dipole and Casimir-Polder interactions.
However, as pointed out in~\cite{vandeKamp:2020rqh,Schut:2023eux}, another way to secure that the gravitational-induced entanglement is dominating is to introduce a grounded conducting plate between the test masses that shields any electromagnetic interactions between the test masses, see fig.~\ref{fig:scheme_setup}. 
A common material for the electromagnetic shielding would be copper~\cite{vandeKamp:2020rqh}, or gold-coated silicon-nitride~\cite{Schut:2023eux}, but one could also use thinner, more exotic materials such as graphene~\cite{pavlou2021effective}.

\begin{figure}[t]
    \centering
    \includegraphics[width=\linewidth]{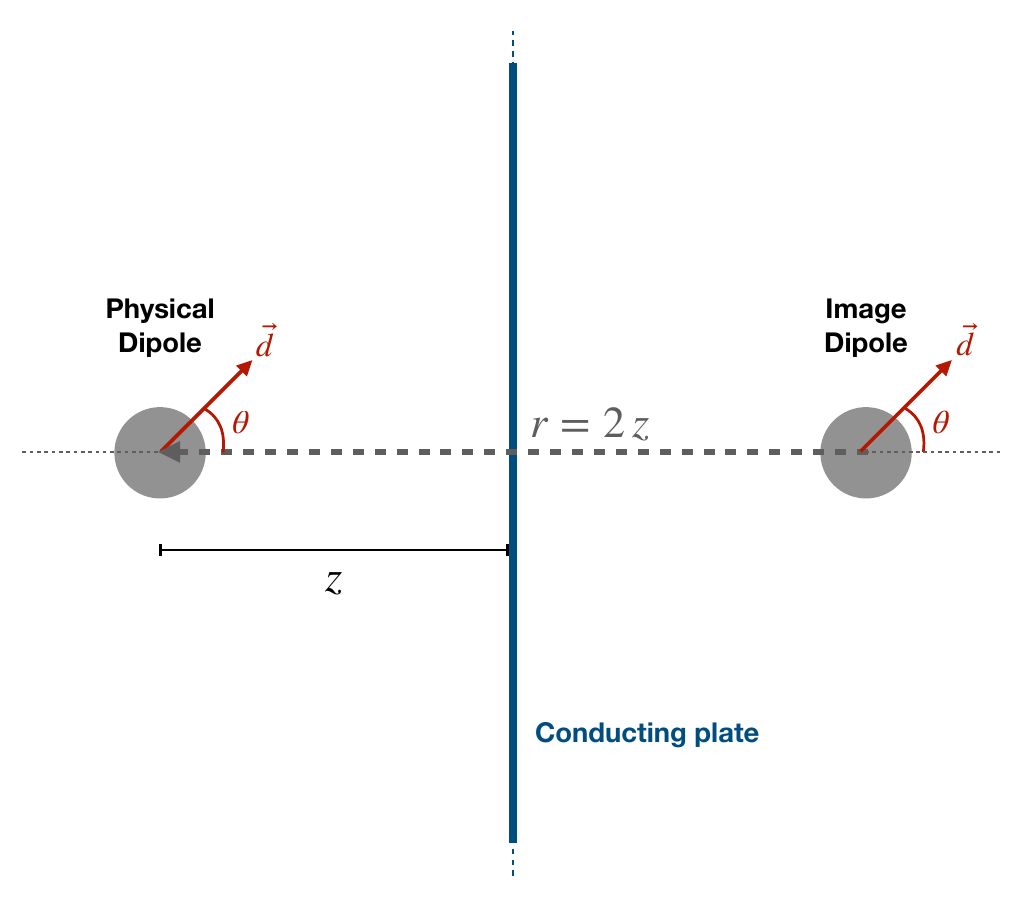}
    \caption{ The figure illustrates the dipole moment of the sphere interacting with the conducting plate. The potential in eq.~\eqref{eq:dip_pot} between the sphere and the plate is found using the method of images~\cite{griffiths2005introduction} from eq.~\eqref{eq:dipole_SS}, where we assume the plate to be grounded and much longer than the radius of the test masses.
    The vector $r$ corresponds to $\boldsymbol{r}$ in eq.~\eqref{eq:dipole_SS}.
    }
    \label{fig:dipole}
\end{figure}

Using electromagnetic shielding we may be able to relax the minimal distance between the test masses. 
Since the gravitational interaction (gravity cannot be screened) is inversely proportional to the separation between the two spheres, decreasing the separation between the spheres increases the gravitational entanglement (see eq.~\eqref{eq:phases}).
An increase in the entanglement means that some of the experimental parameters that were proposed for setups without electromagnetic shielding can be now relaxed (these results are presented in sec.~\ref{sec:results}).
However, the constraint will now be on how close to the plate we can bring the test masses.
Since the shielding plate is conducting, there is still a dipole-induced-dipole and Casimir-Polder interaction between the spherical test mass and the plate (denoted here by S-P).
These interaction potentials are given by ~\cite{Casimir:1948dh,vandeKamp:2020rqh}:~\footnote{For the Casimir-Polder interaction we have assumed that the dielectric properties of the test masses are independent of the frequency of the electric field and we will also assume that its imaginary part is negligible at low temperatures (see experimental findings in~\cite{Floch2011:EPO,Ye2005:die,Ibaraa1997:wide}).
For the dipole sphere-plate potential we use the image dipole procedure~\cite{griffiths2005introduction}, see the schematic explanation in fig.~\ref{fig:dipole}.
}
\begin{align}
    V_\text{DD}^\text{S-P} &=  - \frac{\abs{\boldsymbol{d}}^2}{z^3} \left[ 1+ \cos^2(\theta) \right]\, , \label{eq:dip_pot}\\
    V_\text{CP}^\text{S-P} &= - \frac{3\hbar c}{8\pi} \frac{\varepsilon-1}{\varepsilon+2} \frac{R^3}{z^4} \, , \label{eq:cp_pot}
\end{align}
where $\theta$ is the angle between the vector going from the plate to the test mass, $z$ is the distance between the plate and the test mass (see the schematic of fig~\ref{fig:dipole}), and the magnitude of the electric dipole moment is denoted $\abs{\boldsymbol{d}}$. 

Due to these electromagnetic interactions, the spheres will move towards or away from the plate, depending on the orientation and magnitude of the sphere's dipole moment. 
Since any collision with the plate should be averted, one either needs to very carefully know the initial condition of the test masses (such as dipole moment orientation) and find the initial distance such that during the time of the experiment there is no collision with the plate, or one needs to trap the test masses in the direction towards the plate.
The first setup of test masses that are free to move in the direction perpendicular to the plane of the plate was examined in~\cite{Schut:2023eux}.
The second situation of trapped and shielded test masses will be considered here. 
A schematic setup is shown in fig.~\ref{fig:scheme_setup}.


\subsection{Magnetic trapping}\label{subsec:trapping}

We now analyze the needed characteristics of the magnetic trapping needed to prevent the test masses from colliding with the plate~\footnote{
In the presence of a magnetic field besides eqs.~\eqref{eq:casimir_SS}\eqref{eq:dipole_SS}, there is also an induced magnetic dipole-dipole potential between the two diamagnetic spheres given by
$U_\text{dd}=2\chi_{\rho}^2 m^2 \abs{B^2}/(4\pi \mu_0 r^3)$, where $\chi_\rho=-6.2 \times 10^{-9} {\rm m^3/kg}$ is the mass susceptibility for the diamond-type crystal, $\mu_0$ is the vacuum permeability, $m$ is the mass of the diamonds, and $r$ is their separation.
Being an induced dipole effect, this potential varies quadratically with an applied magnetic field. 
Although this induced magnetic dipole does not significantly interact with the conducting plate, the dipole-induced magnetic potential can also entangle the two masses, see~\cite{Marshman:2023nkh}. 
However, for the magnetic field $\abs{B}\sim {\cal O}(10-50){\rm \mu T}$ and for the separation of $d \sim {\cal O}(20-60){\rm \mu m}$, the induced magnetic dipole potential remains subdominant compared to the gravitational potential between the two masses.
\label{footnote:induced_magnetic_dipole}}.
For a diamagnetic induced magnetic trapping by the trapping magnetic field profile $\boldsymbol{B}_\text{T}$, the trapping potential is given by:
\begin{equation}
    V_\text{T} = \frac{\chi_\rho m \left(\boldsymbol{B}_\text{T}\right)^2}{2\mu_0}\, , \label{eq:trapping_pot}
\end{equation}
with $\chi_\rho$ as the mass magnetic susceptibility of diamond in our case and $\mu_0$ as the vacuum magnetic permeability.
We require the trap to limit the movement in the direction of the plate, e.g. $z$-direction, see fig.~\ref{fig:scheme_setup}. 
Therefore, we require that the trapping potential and the trapping force should be bigger than the dipole-dipole and the Casimir-Polder potentials and the forces along the $\hat{z}$-direction (the direction perpendicular to the plane of the plate):
\begin{equation}\label{Cond-1}
V_\text{DD}+V_\text{CP} < V_\text{T} \, , \qq{} F_\text{DD} + F_\text{CP} < F_\text{T} \, .
\end{equation}
These conditions put constraints on the magnetic field magnitude, and the gradient in the z-direction:~\footnote{
The force is found from $\boldsymbol{F} = -\partial V / \partial \boldsymbol{r}$, for each of the potentials in eqs.~\eqref{eq:dip_pot},~\eqref{eq:cp_pot},~\eqref{eq:trapping_pot}:
\begin{align}
    \boldsymbol{F}_\text{DD} &=  - \frac{3 \abs{\boldsymbol{d}}^2}{z^4} \left[ 1+ \cos^2(\theta) \right] \hat{z}\, , \\
    \boldsymbol{F}_\text{CP} &= - \frac{3\hbar c}{2\pi} \frac{\varepsilon-1}{\varepsilon+2} \frac{R^3}{z^5} \hat{z} \, , \\
    \boldsymbol{F}_\text{T} &= \frac{2 \chi_\rho m}{2\mu_0} \left( \boldsymbol{B}_\text{T}\pdv{\boldsymbol{B}_\text{T}}{x} \hat{x} + \boldsymbol{B}_\text{T} \pdv{\boldsymbol{B}_\text{T}}{y} \hat{y} + \boldsymbol{B}_\text{T} \pdv{\boldsymbol{B}_\text{T}}{z} \hat{z} \right) \, .
\end{align}
The Casimir-Polder~\cite{Ford:1998ex,vandeKamp:2020rqh} and the dipole-dipole interactions are assumed to be only in the $\hat{z}$-direction.
For the magnetic force, only the $\hat{z}$-component is considered in eq.~\eqref{eq:force_ineq}.
Furthermore, $\boldsymbol{B}_\text{T} \cdot \pdv{\boldsymbol{B}_\text{T}}{z} = \abs{\boldsymbol{B}_\text{T}} \abs{\partial_z \boldsymbol{B}_\text{T}} \cos(\phi) \leq \abs{\boldsymbol{B}_\text{T}}\abs{\partial_z \boldsymbol{B}_\text{T}}$ is used to find eq.~\eqref{eq:force_ineq}}
\begin{align}
    \abs{\boldsymbol{B}_\text{T}} > \left[\frac{2\mu_0}{\chi_\rho m} (V_\text{DD} + V_\text{CP} ) \right]^{1/2} \, , \label{eq:pot_ineq} \\
    \abs{\partial_z \boldsymbol{B}_\text{T}} > \frac{\mu_0}{\chi_\rho m \abs{\boldsymbol{B}_\text{T}}} (F_\text{DD} + F_\text{CP} )\, . \label{eq:force_ineq}
\end{align}
For a diamond test mass, we have taken the dielectric constant $\epsilon=5.1$, the density $\rho=3500\,\si{\kilogram\per\metre\cubed}$, and the magnetic susceptibility $\chi_\rho = - 6.2 \times 10^{-9} \,\si{\metre\cubed\per\kilogram}$.
The electric dipole moment has two contributions, a permanent and the induced one. The permanent electric dipole moment magnitude has been measured in~\cite{Afek:2021bua} for a silica type crystal with radii $5-10\,\si{\micro\metre}$, and was found to be $\sim10^{-4}\,e \,\si{\metre}$ (with $e$ the electric charge unit), and to have no significant mass-dependence. 
Studies on other microspheres~\cite{Rivic} indicate that there could be a volume scaling for the permanent dipole moment.
In the context of the QGEM experiment we consider diamond spheres of mass $10^{-16}-10^{-13}\,\si{\kilogram}$, which would correspond to $R\approx 0.2-2\,\si{\micro\metre}$.
Therefore, here we assume a permanent dipole moment of $\sim10^{-4}\,e \,\si{\metre}$ that is independent of the radius of the sphere.

The induced electric dipole moment's magnitude is given by 
$4\pi\epsilon_0\alpha E$, where $E$ is the external electric field that induces the dipole and $\alpha$ is the polarisability of the diamond, given by $R^3(\varepsilon-1)/(\varepsilon+2)$ (see footnote~\ref{footnote:cp}).
We can consider for example the electric field from a wire. 
The wire could be used to create superpositions~\cite{Zhou:2022frl}.
The ampacity for copper nanotubes is $J=10^{13}\,\si{\ampere\per\metre\squared}$~\cite{Subramaniam2013:one}, with a thermal conductivity of $\sigma = 4.6\times10^{7} \,\si{\siemens\per\meter}$ at a room temperature~\cite{Subramaniam2013:one}, the electric field found from Ohm's law is: $E = J/\sigma \sim 2\times10^{5} \,\si{\metre\kilogram\per\second\cubed\per\ampere}$.
Therefore, the magnitude of the corresponding induced dipole moment for  test masses $10^{-16}-10^{-13}\,\si{\kilogram}$ would be
$ 5\times 10^{-4} - 5\times 10^{-7}\,e\,\si{m}$.

For our present analysis, we will assume that the test masses are well-isolated and that there is a negligible external electric field,  we  will assume here that the magnitude of the electric dipole moment is mass-independent and it is given by $\abs{\boldsymbol{d}}\sim10^{-4}\,e \,\si{\metre}$.
Furthermore, we assume a maximal effect from the dipole interaction by taking the angle between the direction of the plate to the center of mass of the diamond sphere and the electric dipole moment vector to be $\theta=2\pi k$, $k\in\mathbb{Z}$.


\begin{figure}[t]
    \centering
    \includegraphics[width=\linewidth]{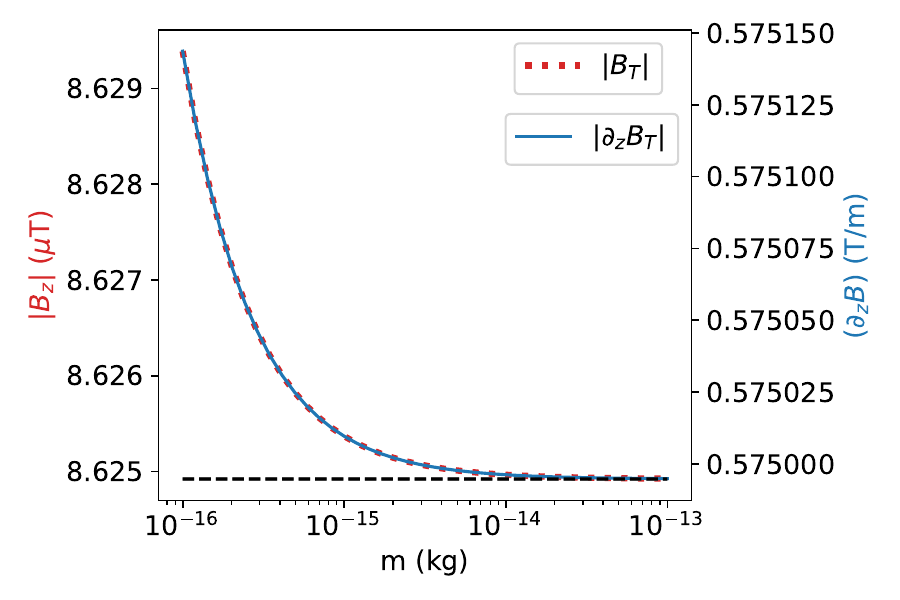}
    \caption{ The plot shows the minimal magnetic field strength (dotted, red) and its gradient (solid, blue) along the z-direction (see fig.~\ref{fig:scheme_setup} for the geometric setup) as a function of the mass for a diamond sphere, where the trapping potential overcomes the Casimir-Polder and the dipole interaction in the $\hat{z}$-direction. The plate  is located at a fixed distance ($30\,\si{\micro\metre}$) from the diamond, and we consider the dipole moment of the diamond to be $\abs{\boldsymbol{d}} = 10^{-4}\,e\,\si{\metre}$ (and the dipole orientation with regards to the plate is $\theta=0$). Here, we consider the mass of the diamond to be $10^{-14}$kg with a radius of $0.88{\rm \mu m}$.
    The dashed black line indicates the magnetic field strength and its gradient along the $z$-direction when the dipole moment is assumed to have a volume scaling, where a sphere of radius $10\,\si{\micro\metre}$ with $\abs{\boldsymbol{d}}=10^{-4}\,e\,\si{\metre}$ is taken as a baseline~\cite{Afek:2021bua}.
    }
    \label{fig:trap_mass}
\end{figure}


\begin{figure*}[t]
\includegraphics[width=0.49\linewidth]{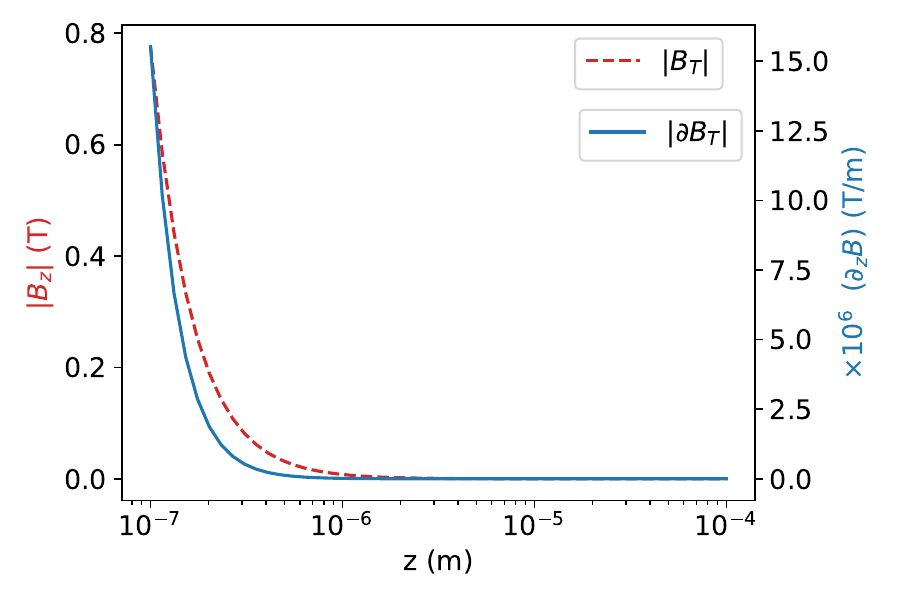}
\includegraphics[width=0.49\linewidth]{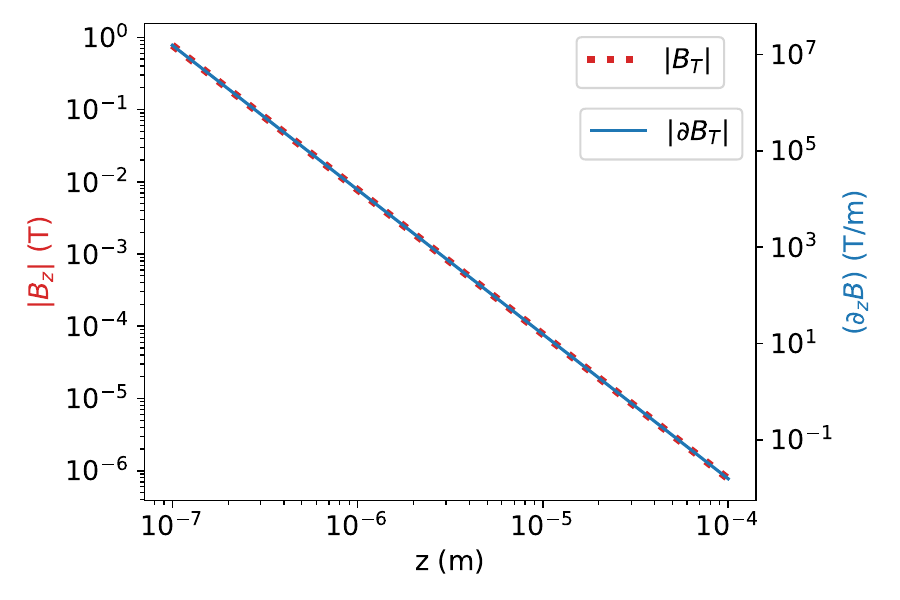}
\caption{In this plot we show the minimal magnetic field strength (dashed, red) and gradient (solid, blue) where the trapping potential overcomes the Casimir-Polder and dipole interaction towards the plate (see eqs.~\eqref{eq:pot_ineq},~\eqref{eq:force_ineq} and eqs.~\eqref{eq:dip_pot},~\eqref{eq:cp_pot},~\eqref{eq:trapping_pot}), as a function of the separation of a diamond sphere and the conducting plate. 
We plot for two different scales on the $y$-axis. The left plot with linear scale shows the rapid increase for $z<10^{-6}\,\si{\metre}$. The right plot with log-scale $y$-axis shows more clearly the numerical values of $\abs{\boldsymbol{B_z}}$ and $\abs{\partial_z\boldsymbol{B}}$ as a function fo $z$.
For a test mass of $10^{-14}\,\si{\kilogram}$ and diamond spheres with an electric dipole moment of $\abs{\boldsymbol{d}} = 10^{-4}\,e\,\si{\metre}$ (and $\theta=0$).}
\label{fig:trap_constraints}
\end{figure*}


The minimal values for the magnetic field strength and the magnetic field gradient in the $\hat{z}$-direction should be such that there is no net force towards the plate, the conditions in eq.~\ref{Cond-1} will then yield the plots for the magnetic field $B_{\rm T}$ and $\partial_z B_{\rm T}$. 
These conditions are plotted with respect to the mass of the diamond, see fig.~\ref{fig:trap_mass}, and with respect to the diamond-plate separation, see fig.~\ref{fig:trap_constraints}.

Since the dipole moment is taken to be mass-independent, the magnetic field strength scales inversely with the mass (see eqs.~\eqref{eq:pot_ineq},~\eqref{eq:force_ineq} and eqs.~\eqref{eq:dip_pot},~\eqref{eq:cp_pot},~\eqref{eq:trapping_pot}).
The dashed-black line in fig.~\ref{fig:trap_mass} also shows the requirement for a dipole moment that scales with volume, in which case the requirement on the magnetic field strength becomes mass-independent.
From fig.~\ref{fig:trap_mass}, we can see that for the range of diamond masses, relevant to the QGEM experiment, e.g., $10^{-16}-10^{-13}\,\si{\kilogram}$, kept at a distance of $30\,\si{\micro\metre}$ from the plate, we  require $\abs{\boldsymbol{B}_\text{T}}>8.7\,\si{\micro\tesla}$ and $\abs{\partial_z\boldsymbol{B}_\text{T}}>0.58\,\si{\tesla\per\metre}$, an experimentally reasonable magnetic field and gradient along the $\hat z$-direction.

The bounds on the magnetic field strength and gradient are independent of the details of the trapping profile.
However, they are dependent on the separation between the test mass and the plate, which is partly fixed by the trapping profile in the sense that a `flatter' trap may require a larger separation. 
Figure~\ref{fig:trap_constraints} shows the minimal requirement on the magnetic field magnitude and the gradient as a function of the separation between the test mass and the plate. We can see that for the separations $z<1\,\si{\micro\metre}$, the requirements on the magnetic field increase rapidly.
Up to separations of $10\,\si{\micro\metre}$ we would require $\abs{\boldsymbol{B}_z} \sim 10\,\si{\micro\tesla}$ and $\abs{\partial_z\boldsymbol{B}_z} \sim 10\, \si{\tesla/\metre}$. 

For the rest of our computation, we will mainly consider the cases:
$z=30\,\si{\micro\metre}$ and
$z=10\,\si{\micro\metre}$.
From the previous calculation, we find the conditions on the magnetic field and the gradient of the magnetic field along the $\hat z$-direction for masses in the range $10^{-18}-10^{-13}\,\si{\kilogram}$:
\begin{align}\label{eq:fin_conditions}
    \abs{\boldsymbol{B}_\text{T}} > 8.7 \,\si{\micro\tesla} \, , \qq{}
    \abs{\partial_z \boldsymbol{B}_\text{T}} > 0.58 \,\si{\tesla\per\metre} \, ,
\end{align}
for $z=30\,\si{\micro\metre}$ separation from the plate and 
\begin{align}\label{eq:fin_conditions2}
    \abs{\boldsymbol{B}_T} > 78 \,\si{\micro\tesla} \, , \qq{}
    \abs{\partial_z \boldsymbol{B}_T} > 16 \,\si{\tesla\per\metre} \, ,
\end{align}
for $z=10\,\si{\micro\metre}$ separation from the plate (which is also effectively mass-independent for $m\in[10^{-18}-10^{-13}]\,\si{\kilogram}$).

Both the above-mentioned cases are satisfied by the trapping profile created in Ref.~\cite{Hsu:2016} (given in eq.~\eqref{eq:trapping_profile}). 
The authors of Ref.~\cite{Hsu:2016} created a magnetic trap for the feedback cooling of a nanodiamond, they simulated the levitation of the nanocrystal by using its weak diamagnetism in combination with a large magnetic field gradient to counter the gravitational potential.
Using the aforementioned profile, we can for example trap the particle in the $\hat{y}, \hat{z}$-direction, while allowing it to move freely in the $\hat{x}$-direction.
The profile of Ref.~\cite{Hsu:2016} was used in Ref.~\cite{Marshman:2023nkh} to propose a scheme for the creation of a macroscopic quantum superposition within the trap.



\section{Results}\label{sec:results}

To witness the quantum nature of gravity, in the parallel setup, we can construct a witness $\mathcal{W}$, for which we choose the Positive Partial Transpose (PPT) witness~\cite{Chevalier:2020uvv,Tilly:2021qef,Schut:2021svd}.
The expectation value of the PPT witness is given as:
\begin{equation}
    \langle \mathcal{W} \rangle 
    \equiv \Tr(\mathcal{W}\rho^{T_2}) = \Tr(\ket{\lambda_-^{T_2}}\bra{\lambda_-^{T_2}}\rho^{T_2}) = \lambda_-^{T_2} \, , \label{eq:ppt_def}
\end{equation}
where $\lambda_-^{T_2}$ denotes the most negative eigenvalue of the partial transpose of the density matrix, which is denoted $\rho^{T_2}$.
From sec.~\ref{sec:setup}, we already know the most negative eigenvalue, see eq.~\eqref{eq:eigval1}.
However, now we also wish to  include the decoherence rate in the density matrix. We will follow the convention of modelling the decoherence rate as an exponential decay in the density matrix, see Ref.~\cite{Schlosshauer:2019ewh}. The decoherence rate can be added as follows in the density matrix:
\begin{equation}
    \bra{i\,j}\rho\ket{i'\,j'} \rightarrow e^{-\gamma \tau (2 - \delta_{i,i'} - \delta_{j,j'})} \, ,
\end{equation}
with $\gamma$ the decoherence rate, and the bra and ket with $i,j,i',j'=\, \uparrow,\downarrow$ select the elements of the density matrix elements.

By including the decoherence rate in our density matrix, the expectation value of the witness, e.g. the most negative eigenvalue, can thus be expressed as:~\cite{Chevalier:2020uvv,Schut:2023eux}
\begin{align}
    \langle \mathcal{W} \rangle 
    &= \frac{1}{4} - \frac{1}{4} e^{-\gamma \tau} \left[ e^{-\gamma \tau} \mp 2 \sin(\frac{\phi_1+\phi_2}{2})\right] \, .\label{eq:witness_def}
\end{align}
Note that in eq.~\eqref{eq:witness_def}, when $\gamma=0$ we obtain the expression of the most negative eigenvalue of the partial transpose density matrix given by eq.~\eqref{eq:eigval1} in the absence of decoherence.
In appendix~\ref{app:witness} we give the full derivation of the witness in terms of it's minimal eigenvectors and expansion in terms of the Pauli matrices.

Due to $\phi_1+\phi_2$ being negative for the parallel setup, the most negative eigenvalue in eq.~\eqref{eq:witness_def} is with the `$+$' sign. In contrast, for the linear setup, the argument of the sin function is positive, and the most negative eigenvalue is with the `$-$' sign (see the discussion in sec.~\ref{sec:setup}).
The expectation value of the witness can be approximated for small small-time intervals from eq.~\eqref{eq:witness_def}, writing in a setup-independent (parallel or linear) manner gives:
\begin{equation}
    \langle \mathcal{W} \rangle \approx \frac{1}{2} \gamma \tau - \frac{1}{2} \abs{\omega_\text{ent}} \tau \, , \label{eq:witness_approx}
\end{equation}
where $\gamma$ is the decoherence rate and we define the entanglement rate as 
\begin{equation}
\omega_\text{ent} \, \tau \equiv \phi_\text{ent} \equiv (\phi_1+\phi_2)/2 \, \label{eq:omega_ent}
\end{equation}
The absolute value is because in the parallel setup, this value is negative and we would have $\frac{1}{2}(\gamma + \omega_\text{ent})\tau$, while for the linear setup, the entanglement rate is positive, giving $\frac{1}{2}(\gamma - \omega_\text{ent})\tau$.
We use the absolute sign to write the expectation value of the witness in a more generic way that is applicable to any 2-qubit setup.
We refer to appendix~\ref{app:witness} for a more detailed derivation.

\begin{figure}[t]
    \centering
    \includegraphics[width=\linewidth]{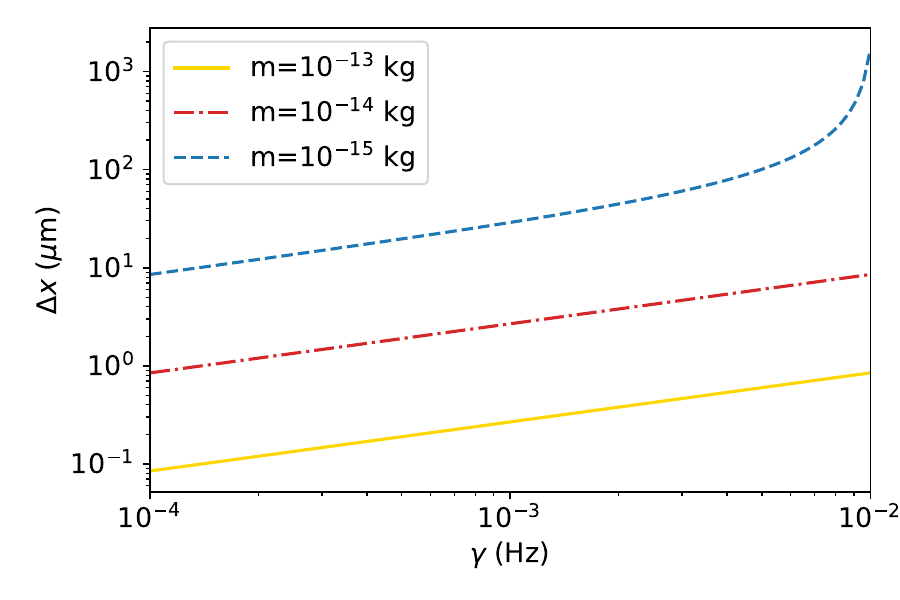}
    \caption{The plot shows the superposition width $\Delta x$ as a function of the decoherence rate $\gamma$. The lines indicate the value of $\Delta x$ such that $\langle\mathcal{W}\rangle =0$, the area above these lines are the values for $\Delta x$ for which $\langle\mathcal{W}\rangle <0$, hence, the entanglement can be witnessed. 
    We have chosen the value of  $d=61\,\si{\micro\metre}$ (the distance between the two superpositions), which corresponds to a sphere-plate separation of $z=30\,\si{\micro\metre}$ and $1\,\si{\micro\metre}$ for the conducting plate's thickness.}
    \label{fig:decoherence}
\end{figure}

For the parallel setup, the entanglement rate is given by:
\begin{equation}\label{eq:phi_ent}
\omega_\text{ent} \equiv \phi_1^\text{par} / \tau
= \left[\frac{1}{\sqrt{d^2+(\Delta x)^2}} - \frac{1}{d} \right] \frac{Gm^2}{\hbar} \, , 
\end{equation}
where $\phi_1^\text{par}$ is given by eq.~\eqref{eq:phases}.
If $\Tr(\mathcal{W}\rho)<0$ then the gravity-mediated entanglement can be detected. This gives us a relation between the allowed decoherence rate $\gamma$, and the entanglement rate, $\omega_\text{ent}$:
\begin{equation}\label{eq:witness_ineq}
    \text{if } \gamma < \abs{\omega_\text{ent}} \Rightarrow \text{ entanglement can be witnessed}\, .
\end{equation}
We can find the minimal required superposition width $\Delta x$ based on the expected decoherence rate from the witness in eq.~\eqref{eq:witness_approx}, as:
\begin{equation}
    \Delta x > \sqrt{\left(\frac{G m^2}{G m^2 / d - \gamma \hbar + \langle \mathcal{W}\rangle \hbar/2\tau}\right)^2-d^2} \, .
\end{equation}
This equation is plotted in fig.~\ref{fig:decoherence} as a function of the decoherence rate $\gamma$, for a witness value of $\langle\mathcal{W}\rangle=0$ and a separation of $d=61\,\si{\micro\metre}$ (which would correspond to a sphere-plate separation of $z=30\,\si{\micro\metre}$ and a $1\,\si{\micro\metre}$ thick plate).
The figures show that a larger superposition width is required for a larger decoherence rate to find  a negative expectation value of the  witness, hence the entanglement.

It should be noted that we have made a few simplifications.
In general, both $\gamma$ and $\omega_\text{ent}$ are time-dependent since they are a function of $\Delta x$ for a one-loop interferometer.
However, we have considered them to be constant during the time $\tau$.
The time dependence of $\Delta x$ is determined by the exact splitting procedure.
After fixing the splitting procedure (see for example~\cite{Zhou:2022epb,Zhou:2022frl,Toros:2020dbf}) one can perform a more detailed time-dependent analysis of the accumulated phase, decoherence rate, and the witness.
To keep the discussion general, we have neglected the time dependence. 
This is not a bad approximation because if we were to consider the superposition to be created and closed by the pulsed magnetic field during time $\tau_a$ that is smaller than the interaction time $\tau$, during which the superposition can be treated nearly constant (as in~\cite{Bose:2017nin}).
During the time $\tau_a$ the entanglement rate and decoherence rate are smaller because $\Delta x$ is smaller, therefore, the time-independent analysis done here works quite well.
    
If the entanglement rate and the decoherence rate are very close to each other, then the expectation value of the witness will be small, which means that a lot of repetition of the spin measurements will be required to detect the entanglement. 
Hence, although we presented the minimal superposition width to witness the expectation value to be negative if we required the entanglement to be detectable within some finite number of measurements, we would require a larger $\Delta x$ for a lower number of measurements.
Finding the exact $\Delta x$ requirement is based on some set of a number of measurements that would require a numerical simulation of the experiment, such as performed in~\cite{Tilly:2021qef,Schut:2021svd}.

Figure~\ref{fig:decoherence} (which plots the minimal $\Delta x$ as a function of $\gamma$ for $d=61$ micron) shows that the requirements on $\Delta x$ relax for larger masses since the gravitational interaction strength increases.
However, it is also more difficult to create a superposition for a larger mass.
We have noted that the magnetic trapping profile discussed in Ref.~\cite{Marshman:2023nkh} suggests that we could potentially take a separation of $21\,\si{\micro\metre}$ (which would correspond to a sphere-plate separation of $z=10\,\si{\micro\metre}$ and a $1\,\si{\micro\metre}$ thick plate). 
In table~\ref{table:deltax}, we have given an overview of the superposition width $\Delta x$ for different masses and for different values of $z$ (see fig.~\ref{fig:dipole}), such that $\abs{\omega_\text{ent}}=0.01\,\si{\hertz}$ in the parallel setup.

The values in table~\ref{table:deltax} correspond to the expectation value of the witness, $\langle\mathcal{W}\rangle=0$, for a  given decoherence rate of $\gamma\sim\order{10^{-2}}\,\si{\hertz}$.  We have selected the decoherence rate of $\gamma \sim10^{-2}\,\si{\hertz}$ from our previous papers, which  seems reasonable for the ambient temperatures of $\sim 1\,\si{\kelvin}$, based on scattering with air molecules and black body radiation,~\cite{vandeKamp:2020rqh,Chevalier:2020uvv,Schut:2021svd,Tilly:2021qef,Sinha:2022snc}, and dipole-dipole interactions~\cite{Fragolino:2023agd}.
Based on eq.~\eqref{eq:witness_ineq} this would inevitably place a lower bound on the rate of entanglement to be at least $\abs{\omega_\text{ent}} > 10^{-2}\,\si{\hertz}$ and the values of $\Delta x$ in table~\ref{table:deltax} are the minimal values for detecting entanglement. 
We can see the remarkable change in the size of the spatial superposition if we bring the two masses close, to $d= 21 \,{\rm\mu m}$ ($z=10\,{\rm \mu m}$ from the plate) and $d=11\,{\rm \mu m}$ ($z=5\,{\mu m}$ from the plate). 
For $m\sim 10^{-14}$ kg, the spatial superposition reduces to $\Delta x = 1.7 \,{\rm \mu m}$ and $0.65 \,{\rm \mu m}$, respectively. 
The extent that which we can reduce $\Delta x$ for such heavy masses will have a  huge advantage for the QGEM experiment~\footnote{
We want to consider the $z=5{\rm \mu m}$ case further as it requires a detailed study of both the trapping magnetic field profile and the magnetic field profile due to the SG apparatus. 
Furthermore, the induced magnetic dipole interaction mentioned in footnote~\ref{footnote:induced_magnetic_dipole} will be a constraining factor at distances $z=5{\rm \mu m}$ and $z=10{\rm \mu m}$ if we cannot screen the magnetic field as well. 
However, if we were to screen the magnetic field via a superconducting screen, this would also induce a force between the micro-diamond and the screen due to the image magnetic dipole, which requires further investigation.
}. 

Additionally, the plate will be a source of dephasing, this was analysed in Ref.~\cite{Schut:2023eux}.
It was found that the dephasing was mostly due to the permanent dipole of the diamond (the Casimir-Polder interaction with the plate and thermal fluctuations of the plate are sub-dominant sources of dephasing).
Ref.~\cite{Schut:2023eux} showed that in the worst-case scenario, restricting the dephasing to $10^{-1}$ requires sub-femtometer precision on the position of the superposition instances, showing that a good control of the dipole is necessary to reduce the dephasing.
Furthermore, precision of the initial conditions was analysed in Ref.~\cite{Schut:2023eux}.
They estimated a precision of the order of femtometers is necessary to reduce phase fluctuations in repeated measurements (in the worst-case scenario).
These results translate approximately to the setup proposed in this paper, this number can be reduced by mitigating the dipole of the diamond test mass.
 

\begin{table}[t]
\bgroup
\def\arraystretch{1.5}
\begin{tabular}{|| c | c | c | c ||} 
 \hline
 \hfill Mass (kg) \hfill & \hfill \begin{tabular}{@{}c@{}} $\Delta x\, (\si{\micro\metre})$ \\ $z = 30 \,\si{\micro\metre}$ \end{tabular} \hfill & 
 \hfill \begin{tabular}{@{}c@{}} $\Delta x\, (\si{\micro\metre})$ \\ $z = 10 \,\si{\micro\metre}$ \end{tabular} \hfill & \hfill \begin{tabular}{@{}c@{}} $\Delta x\, (\si{\micro\metre})$ \\ $z = 5 \,\si{\micro\metre}$ \end{tabular} \hfill \\ [0.5ex] 
 \hline\hline
 \hfill $10^{-15}$ \hfill & \hfill $1685$ \hfill & \hfill $23$ \hfill & \hfill $7.5$ \hfill
 \\ 
 \hline
 \hfill$10^{-14}$ \hfill & \hfill $8.5$ \hfill & \hfill $1.7$ \hfill & \hfill $0.65$ \hfill
 \\ 
 \hline
 \hfill$10^{-13}$ \hfill & \hfill $0.85$ \hfill & \hfill $0.17$ \hfill & \hfill $0.06$ \hfill
 \\  
 \hline
\end{tabular}
\egroup
\caption{The table shows the mass and the superposition width ($\Delta x$) required to get the desired entanglement rate of $\abs{\omega_\text{ent}} = 0.01 \,\si{\hertz}$, see eq.~\eqref{eq:phi_ent}. 
For two test masses in a quantum superposition in their magnetically trapped potentials, and shielded by a conducting plate of thickness $1{\rm \mu m}$. 
The masses are kept at a distance $30\,\si{\micro\metre}$ (second column), $10\,\si{\micro\metre}$ (third column) and $5\,\si{\micro\metre}$ (fourth column) from the plate. 
Hence, the superposition of the masses is separated by the distance of $d=61\,\si{\micro\metre}$, $d=21\,\si{\micro\metre}$ and $11\,\si{\micro\metre}$, respectively.
For an entanglement rate $>0.01\,\si{\hertz}$, the expectation value of the witness becomes negative, see eq~\eqref{eq:witness_ineq}.
}
\label{table:deltax}
\end{table}

\section{Conclusion \& Discussion}\label{sec:conc}

The electromagnetic screening between the test masses in a parallel configuration in the QGEM protocol has been known to relax the experimental constraints, see ~\cite{Schut:2023eux}. 
However, the experimental parameters did not improve that much compared to the linear setup investigated in~\cite{vandeKamp:2020rqh}. This is because one has to duly consider the Casimir-Polder and the dipole interactions between the test mass and the conducting plate. A recent analysis of Ref.~\cite{Schut:2023eux} suggested that for mass of the superposition around $10^{-14}$ kg, we would still require the superposition size to be within $\Delta x \sim {\cal O}(10-20){\rm \mu m}$. A similar bound was also obtained in~\cite{vandeKamp:2020rqh} for the linear setup. Since creating the massive spatial superposition is still one of the main challenges for the QGEM experiment, if this requirement of $\Delta x$ becomes less stringent then it will indeed be a great improvement for the practical aspects of the experiment.

In this paper, we have shown that a combination of electromagnetic screening and the magnetic trapping of the test mass can optimize the experimental parameters in the QGEM protocol. 
We have shown that the trapping of the test masses constricts the trajectories in such a way that the superposition of the test masses is now created parallel to the conducting plate. 
Furthermore, the trapping potential forbids the particles to come very close to the conducting plate as well. This is possible due to the fact that the trapping potential can provide a flat direction, where the potential is flat (or weakly confined) and along this flat direction the superposition can be created. 
One such example has been studied recently in \cite{Marshman:2023nkh}.

We have found that the combination of screening and trapping reduces the requirements for the superposition size.
For a mass of $10^{-14}\,\si{\kilogram}$ we would now require the maximum superposition width of approximately $3.4\,\si{\micro\metre}$ for an entanglement phase of $0.01$ rad in $1\,\si{\second}$, as opposed to $\Delta x \sim 100\,{\rm \mu m}$ for the same entanglement rate. This is roughly two orders of magnitude improvement compared to the setup without the conducting plate~\cite{Bose:2017nin}, and an order of magnitude improvement compared to the shielded free-fall setup~\cite{Schut:2021svd,vandeKamp:2020rqh}. 

The next task will be to study the common mode fluctuations of the conducting plate in the presence of the trapping potential and study various sources of decoherence. We intend to do this in the future paper. The trapping and the shielding protocol are extremely promising in the realizations of the QGEM experiment. Therefore, we should perform a more detailed analysis to find various electromagnetic and mechanical sources of decoherence. Furthermore, other sources of noise such as gravity gradient noise and relative acceleration noise has to be re-evaluated based on our earlier analysis~\cite{Toros:2020dbf}.

Since the most challenging aspect of all is to create a massive and spatial splitting of the wave function, here we have shown that the QGEM experiment can be performed for a tiny splitting of the wave function for a heavy massive object, as shown in table~\ref{table:deltax}, if we can simultaneously  trap and shield the two matter-wave interferometers. Indeed, creating the initial state will be a common challenge for any mechanisms for creating the superposition, see~\cite{Deli__2020,Tebbenjohanns_2020}, which requires to be addressed within this trapping potential, and beyond the scope of the current analysis.

\section*{Acknowledgements} \label{sec:acknowledgements}
MS is supported by the Fundamentals of the Universe Fellowship within the University of Groningen.
A.G. is supported in part by NSF grants PHY-2110524 and PHY-2111544, the Heising- Simons Foundation, the John Templeton Foundation, the W. M. Keck Foundation, and ONR Grant N00014-18- 1-2370. 
S.B. would like to acknowledge EPSRC grants (EP/N031105/1, EP/S000267/1, and EP/X009467/1) and grant ST/W006227/1.

\appendix
\section*{Appendix A: witness} \label{app:witness}
The expectation value of the PPT (positive partial transpose) witness given in eq.~\eqref{eq:ppt_def}:
\begin{equation}
    \langle \mathcal{W} \rangle 
    \equiv \Tr(\mathcal{W}\rho^{T_2}) = \Tr(\ket{\lambda_-^{T_2}}\bra{\lambda_-^{T_2}}\rho^{T_2}) = \lambda_-^{T_2} \, ,
\end{equation}
shows that the witness $\mathcal{W}$ is defines by
\begin{equation}
    \mathcal{W} = \ket{\lambda_-^{T_2}}\bra{\lambda_-^{T_2}} \, , \label{eq:ppt_def2}
\end{equation}
it is the partial transpose of the tensor product of eigenvectors that correspond to the minimal eigenvalues of the partial transpose density matrix. 
Including decoherence as an exponential decay, the density matrix of the wavefunction presented in eq.~\eqref{WF-1} is given by:
\begin{align}\label{denmat}
&\rho = \nonumber \\
&\frac{1}{4}
    \begin{pmatrix}
        1 & e^{(-i\omega_2- \gamma) \tau} & e^{(-i\omega_1 - \gamma )\tau} & e^{- 2\gamma \tau} \\
        e^{(i\omega_2 - \gamma )\tau} & 1 & e^{-i(\omega_1-\omega_2)- 2 \gamma \tau} & e^{(i\omega_2 - \gamma )\tau} \\
        e^{(i\omega_1 - \gamma )\tau} & e^{i(\omega_1-\omega_2)- 2 \gamma \tau} & 1 & e^{(i\omega_1 - \gamma )\tau} \\
        e^{ - 2 \gamma \tau} & e^{(-i\omega_2 - \gamma )\tau} & e^{(-i\omega_1 - \gamma )\tau} & 1
    \end{pmatrix} 
\end{align}
where $\omega_{1,2}\tau=\phi_{1,2}$ and $\phi_1$, $\phi_2$ defined in eqs.~\eqref{eq:phases},~\eqref{eq:phases_lin} for the parallel and linear setup, respectively.
Note that eq.~\eqref{denmat} at $t=0$ is a pure state, we assume that the two matter-wave interferometers can be prepared in an pure state, although this may be experimentally challenging. 
After taking the partial transpose one can find the eigenvalues:
\begin{align}
    \lambda_{1,2} &= \frac{1}{4} - \frac{1}{4} e^{-\gamma \tau} \left[ e^{-\gamma \tau} \mp 2 \sin(\frac{\phi_1+\phi_2}{2})\right] \, , \\
    \lambda_{3,4} &= \frac{1}{4} + \frac{1}{4} e^{-\gamma \tau} \left[ e^{-\gamma \tau} \pm 2 \cos(\frac{\phi_1+\phi_2}{2}) \right]  \, .
\end{align}
For $\gamma=0$, they reduce to the eigenvalues in eqs.~\eqref{eq:eigval1},~\eqref{eq:eigval2}.
The eigenvectors corresponding to the min eigenvalues for the parallel ($\lambda_-^\text{par}=\lambda_1$) and  the linear case ($\lambda_-^\text{lin}=\lambda_2$) are given by:
\begin{align}
    \ket{\lambda_-^\text{lin}} &= (-1, -i, i, 1)^T \, , \\
    \ket{\lambda_-^\text{par}} &= (-1, i, -i, 1)^T \, ,
\end{align}
for $\phi_1=\phi_2=0$, and taking $\ket{\downarrow}\to(1,0)^T$ and $\ket{\uparrow}\to(1,0)^T$.
The witness is constructed from eq.~\eqref{eq:ppt_def2}.
The witness matrix can be expanded in terms of the identity matrix $I$ and products of the Pauli matrices $\sigma_x$, $\sigma_y$ and $\sigma_z$.
The expansion is given by:
\begin{align}
    \text{parallel: } \mathcal{W}_+ &= \frac{1}{4} \left(1 \otimes 1 - X \otimes X + Z \otimes Y + Y \otimes Z \right) \, , \\
    \text{linear: } \mathcal{W}_- &= \frac{1}{4} \left(1 \otimes 1 - X \otimes X - Z \otimes Y - Y \otimes Z \right) \, ,
\end{align}
where the expectation value of the witness for the linear case was derived in Ref.~\cite{Chevalier:2020uvv}, and 
following Ref.~\cite{Chevalier:2020uvv}, and by using the decomposition in terms of the Pauli matrix products, we can find an expression for the expectation value of the witness:
\begin{align}
    \Tr(\mathcal{W}_\pm\rho) 
    &= \Tr(\rho) \mp 2 \Im(\rho_{12}) \mp 2 \Im(\rho_{13})- 2 \Re(\rho_{14}) \nonumber \\ &  - 2 \Re(\rho_{23}) \pm 2 \Im(\rho_{24}) \pm 2 \Im(\rho_{34}) \,.
\end{align}
By using the density matrix presented previously, see eq.~\eqref{denmat}, the expectation value of the witness can be computed:
\begin{align}
    4\langle\mathcal{W_\pm}\rangle = 1 - e^{-\gamma t} \bigg( \mp &\left[\sin(\phi_1)+\sin(\phi_2)\right] \nonumber \\ 
    &+ \frac{1}{2} e^{-\gamma t} \left[1+\cos(\phi_2-\phi_1)\right] \bigg) \, .
\end{align}
By considering the small-time expansion around $t=0$, we can find an approximate expression for the expectation value of the witness,  given by:
\begin{align}
    4 \langle W_\pm \rangle 
    &\approx 1 - (1-\gamma t) \bigg[ \mp (\phi_1 + \phi_2) \\ &\hspace{2.5cm}+ \frac{1}{2}(1-\gamma t)(2-\frac{(\phi_2-\phi_1)^2}{2}) \bigg]
\end{align}
At a linear order in $t$, the expectation value of the witness is given by:
\begin{align}
    \text{parallel: } \mathcal{W}_+ &= \frac{1}{4} (2\gamma + 2\omega_1^\text{par}) t \, , \\
    \text{linear: } \mathcal{W}_- &= \frac{1}{4} (2\gamma - \omega_1^\text{lin}-\omega_2^\text{lin}) t \, ,
\end{align}
where $\omega_{1,2} t \equiv \phi_{1,2}$ is the entanglement generation rate, and $\phi_{1,2}$ for the parallel and the linear setups given by eqs.~\eqref{eq:phases},~\eqref{eq:phases_lin}, respectively.
As discussed in sec.~\ref{sec:setup}, for the parallel case $2\phi_1<0$, while for the linear case $\phi_1+\phi_2>0$, for any $\Delta x>0$.
Therefore the witness for the linear and the parallel case can be given more generically as:
\begin{equation}
    \mathcal{W} = \frac{1}{2}(\gamma - \abs{\omega_\text{ent}}) t
\end{equation}
where $\omega_\text{ent} = (\phi_1+\phi_2)/(2t)$, see eqs.~\eqref{eq:omega_ent},~\eqref{eq:phi_ent}.

\section*{Appendix B: Trapping potential} \label{app:trapping-potential}
The magnetic trapping profile from Ref.~\cite{Hsu:2016} that was used for the feedback cooling of a nanodiamond is given in eq.~\eqref{eq:trapping_profile} below. 
The same profile was used in the scheme to create a macroscopic quantum superposition in Ref.~\cite{Marshman:2023nkh}.
\begin{widetext}
\begin{align}
    \boldsymbol{B}_T = 
    & - \Bigg[ \frac{2a_3\sqrt{\frac{14}{3\pi}}xz}{y_0^2} \Bigg] \hat{x} 
    - \Bigg[ \frac{3a_4\sqrt{\frac{35}{\pi}}z}{16y_0^3} \bigg(z^2-3y^2 \bigg) - \frac{a_3\sqrt{\frac{7}{6\pi}}zy}{y_0^2} + \frac{a_2\sqrt{\frac{15}{\pi}}z}{4y_0} \Bigg] \hat{y} \nonumber \\
    &- \Bigg[ \frac{3a_4\sqrt{\frac{35}{\pi}}y}{16y_0^3} \bigg(3 z^2 - y^2\bigg) + \frac{a_3\sqrt{\frac{7}{6\pi}}}{2 y_0^2} \bigg(4 x^2 - y^2 - 3 z^2 \bigg) + \frac{a_2\sqrt{\frac{15}{\pi}}y}{4y_0} \Bigg] \hat{z} \label{eq:trapping_profile}
\end{align}
\end{widetext}
where the directions $\hat{x}, \hat{y}, \hat{z}$ are as indicated in fig.~\ref{fig:scheme_setup}, $y_0$ is the distance of the bottom of the trap to the magnets, and the coefficients $a_{2,3,4}$ are given in units Tesla, and determine the magnetic field strength.
These coefficients were taken to be $a_2 = -1.3\,\si{\tesla}$, $a_3 = 0.0183\,\si{\tesla}$ and $a_4 = 0.72\,\si{\tesla}$ in~\cite{Hsu:2016,Marshman:2023nkh}.

Time-dependent electric and magnetic fields can cause decohere via the dipole-phonon interaction. 
The proposed gradients and fields are not large compared with what has been accomplished experimentally before, and can be achieved for magnets and coils with reasonable mechanical support. 
The largest magnetic field gradients we are talking about (in eqs.~\eqref{eq:fin_conditions},~\eqref{eq:fin_conditions2}) is roughly $10\,\si{\tesla\per\metre}$, which is being used regularly in labs. 
If the time-dependent electric and magnetic fields do not hit the resonant frequency of the fundamental tone of the phonons, then the decoherence is also very small. 
However, these effects should be considered systematically.

The magnetic field used for creating the superpositions will perturb the trapping magnetic field most strongly in the near-flat direction of the trapping potential.
This perturbing magnetic field is switched on and off to create and recombine the spatial superpositions. 
The switching time will depend on the inductance of coils used to create the magnetic gradients as well as eddy currents, and has been analyzed in Ref.~\cite{Marshman:2021wyk}. 
They give a switching time of $\sim100-160\,\si{\micro\second}$, which we expect that this time will be adequate for the proposed measurements.

In a realistic setup, one also has to consider the practical issue of forces applied to the system when switching magnetic fields. These can in principle be minimized with the use of micro-fabricated coils. For example, for coils of radius 10 $\mu$m, forces of order $10^{-8}$ Newtons would act on the system if such a coil created a 0.01 T field and 100T/m gradient at a distance of $\mu$m for a current of 0.1 Ampere, which is attainable in micron-scale patterned wires \cite{Geraci:atomchip}. A detailed experimental design would need to take such conditions into account and is left for future work.


\bibliography{casimir.bib} 

\begin{thebibliography}{10}

\bibitem{Bose:2017nin}
S.~Bose, A.~Mazumdar, G.~W. Morley, H.~Ulbricht, M.~Toro\v{s}, M.~Paternostro,
  A.~A. Geraci, P.~F. Barker, M.~S. Kim, and G.~Milburn, ``{Spin Entanglement
  Witness for Quantum Gravity},'' {\em Phys. Rev. Lett.}, vol.~119, no.~24,
  p.~240401, 2017.

\bibitem{dyson_is_2013}
F.~Dyson, ``{IS} {A} {GRAVITON} {DETECTABLE}?,'' {\em International Journal of
  Modern Physics A}, vol.~28, p.~1330041, Oct. 2013.

\bibitem{Marletto:2017kzi}
C.~Marletto and V.~Vedral, ``{Gravitationally-induced entanglement between two
  massive particles is sufficient evidence of quantum effects in gravity},''
  {\em Phys. Rev. Lett.}, vol.~119, no.~24, p.~240402, 2017.

\bibitem{ICTS}
\url{https://www.youtube.com/watch?v=0Fv-0k13s_k}, 2016.
\newblock Accessed 1/11/22.

\bibitem{Marshman:2019sne}
R.~J. Marshman, A.~Mazumdar, and S.~Bose, ``{Locality and entanglement in
  table-top testing of the quantum nature of linearized gravity},'' {\em Phys.
  Rev. A}, vol.~101, no.~5, p.~052110, 2020.

\bibitem{Bose:2022uxe}
S.~Bose, A.~Mazumdar, M.~Schut, and M.~Toro\v{s}, ``{Mechanism for the quantum
  natured gravitons to entangle masses},'' {\em Phys. Rev. D}, vol.~105,
  no.~10, p.~106028, 2022.

\bibitem{Danielson:2021egj}
D.~L. Danielson, G.~Satishchandran, and R.~M. Wald, ``{Gravitationally mediated
  entanglement: Newtonian field versus gravitons},'' {\em Phys. Rev. D},
  vol.~105, p.~086001, 2022.

\bibitem{Carney_2019}
D.~Carney, P.~C.~E. Stamp, and J.~M. Taylor, ``Tabletop experiments for quantum
  gravity: a user's manual,'' {\em Class. Quant. Grav.}, vol.~36, p.~034001,
  2019.

\bibitem{Carney:2021vvt}
D.~Carney, ``{Newton, entanglement, and the graviton},'' {\em Phys. Rev. D},
  vol.~105, no.~2, p.~024029, 2022.

\bibitem{Christodoulou:2022vte}
M.~Christodoulou {\em et~al.}, ``{Locally mediated entanglement through gravity
  from first principles},'' 2022.

\bibitem{Vinckers:2023grv}
U.~K.~B. Vinckers, A.~de~la Cruz-Dombriz, and A.~Mazumdar, ``{Quantum
  entanglement of masses with non-local gravitational interaction},'' 3 2023.

\bibitem{Scadron:1991ep}
M.~D. Scadron, {\em {Advanced quantum theory and its applications through
  Feynman diagrams}}.
\newblock 1991.

\bibitem{Donoghue:1994dn}
J.~F. Donoghue, ``{General relativity as an effective field theory: The leading
  quantum corrections},'' {\em Phys. Rev. D}, vol.~50, pp.~3874--3888, 1994.

\bibitem{Donoghue:2012zc}
J.~F. Donoghue, ``{The effective field theory treatment of quantum gravity},''
  {\em AIP Conf. Proc.}, vol.~1483, no.~1, pp.~73--94, 2012.

\bibitem{Hensen:2015ccp}
B.~Hensen {\em et~al.}, ``{Loophole-free Bell inequality violation using
  electron spins separated by 1.3 kilometres},'' {\em Nature}, vol.~526,
  pp.~682--686, 2015.

\bibitem{PhysRevA.46.4413}
A.~Peres, ``Finite violation of a bell inequality for arbitrarily large spin,''
  {\em Phys. Rev. A}, vol.~46, pp.~4413--4414, Oct 1992.

\bibitem{GISIN199215}
N.~Gisin and A.~Peres, ``Maximal violation of bell's inequality for arbitrarily
  large spin,'' {\em Physics Letters A}, vol.~162, no.~1, pp.~15--17, 1992.

\bibitem{Bennett:1996gf}
C.~H. Bennett, D.~P. DiVincenzo, J.~A. Smolin, and W.~K. Wootters, ``{Mixed
  state entanglement and quantum error correction},'' {\em Phys. Rev. A},
  vol.~54, pp.~3824--3851, 1996.

\bibitem{Biswas:2022qto}
D.~Biswas, S.~Bose, A.~Mazumdar, and M.~Toro\v{s}, ``{Gravitational
  Optomechanics: Photon-Matter Entanglement via Graviton Exchange},'' 9 2022.

\bibitem{Bose:2022czr}
S.~Bose, A.~Mazumdar, M.~Schut, and M.~Toro\v{s}, ``{Entanglement Witness for
  the Weak Equivalence Principle},'' {\em Entropy}, vol.~25, no.~3, p.~448,
  2023.

\bibitem{Elahi:2023ozf}
S.~G. Elahi and A.~Mazumdar, ``{Probing massless and massive gravitons via
  entanglement in a warped extra dimension},'' 3 2023.

\bibitem{Hanif:2023fto}
F.~Hanif, D.~Das, J.~Halliwell, D.~Home, A.~Mazumdar, H.~Ulbricht, and S.~Bose,
  ``{Testing Whether Gravity Acts as a Quantum Entity When Measured},'' 7 2023.

\bibitem{vandeKamp:2020rqh}
T.~W. van~de Kamp, R.~J. Marshman, S.~Bose, and A.~Mazumdar, ``{Quantum Gravity
  Witness via Entanglement of Masses: Casimir Screening},'' {\em Phys. Rev. A},
  vol.~102, no.~6, p.~062807, 2020.

\bibitem{Schut:2023eux}
M.~Schut, A.~Grinin, A.~Dana, S.~Bose, A.~Geraci, and A.~Mazumdar,
  ``{Relaxation of experimental parameters in a Quantum-Gravity Induced
  Entanglement of Masses Protocol using electromagnetic screening},'' 7 2023.

\bibitem{Fragolino:2023agd}
P.~Fragolino, M.~Schut, M.~Toro\v{s}, S.~Bose, and A.~Mazumdar, ``{Decoherence
  of a matter-wave interferometer due to dipole-dipole interactions},'' 7 2023.

\bibitem{PhysRevLett.123.083601}
O.~Amit, Y.~Margalit, O.~Dobkowski, Z.~Zhou, Y.~Japha, M.~Zimmermann, M.~A.
  Efremov, F.~A. Narducci, E.~M. Rasel, W.~P. Schleich, and R.~Folman,
  ``${T}^{3}$ stern-gerlach matter-wave interferometer,'' {\em Phys. Rev.
  Lett.}, vol.~123, p.~083601, Aug 2019.

\bibitem{Margalit:2020qcy}
Y.~Margalit {\em et~al.}, ``{Realization of a complete Stern-Gerlach
  interferometer: Towards a test of quantum gravity},'' 11 2020.

\bibitem{Marshman:2021wyk}
R.~J. Marshman, A.~Mazumdar, R.~Folman, and S.~Bose, ``{Constructing
  nano-object quantum superpositions with a Stern-Gerlach interferometer},''
  {\em Phys. Rev. Res.}, vol.~4, no.~2, p.~023087, 2022.

\bibitem{PhysRevLett.125.023602}
J.~S. Pedernales, G.~W. Morley, and M.~B. Plenio, ``Motional dynamical
  decoupling for interferometry with macroscopic particles,'' {\em Phys. Rev.
  Lett.}, vol.~125, p.~023602, Jul 2020.

\bibitem{Zhou:2022frl}
R.~Zhou, R.~J. Marshman, S.~Bose, and A.~Mazumdar, ``{Catapulting towards
  massive and large spatial quantum superposition},'' {\em Phys. Rev. Res.},
  vol.~4, no.~4, p.~043157, 2022.

\bibitem{Zhou:2022jug}
R.~Zhou, R.~J. Marshman, S.~Bose, and A.~Mazumdar, ``{Mass-independent scheme
  for enhancing spatial quantum superpositions},'' {\em Phys. Rev. A},
  vol.~107, no.~3, p.~032212, 2023.

\bibitem{Zhou:2022epb}
R.~Zhou, R.~J. Marshman, S.~Bose, and A.~Mazumdar, ``{Gravito-diamagnetic
  forces for mass independent large spatial quantum superpositions},'' 11 2022.

\bibitem{Marshman:2023nkh}
R.~J. Marshman, S.~Bose, A.~Geraci, and A.~Mazumdar, ``{Entanglement of
  Magnetically Levitated Massive Schr\"odinger Cat States by Induced Dipole
  Interaction},'' 4 2023.

\bibitem{PhysRevA.105.012824}
B.~D. Wood, S.~Bose, and G.~W. Morley, ``Spin dynamical decoupling for
  generating macroscopic superpositions of a free-falling nanodiamond,'' {\em
  Phys. Rev. A}, vol.~105, p.~012824, Jan 2022.

\bibitem{Horodecki:2009zz}
R.~Horodecki, P.~Horodecki, M.~Horodecki, and K.~Horodecki, ``{Quantum
  entanglement},'' {\em Rev. Mod. Phys.}, vol.~81, pp.~865--942, 2009.

\bibitem{Chevalier:2020uvv}
H.~Chevalier, A.~J. Paige, and M.~S. Kim, ``{Witnessing the nonclassical nature
  of gravity in the presence of unknown interactions},'' {\em Phys. Rev. A},
  vol.~102, no.~2, p.~022428, 2020.

\bibitem{Tilly:2021qef}
J.~Tilly, R.~J. Marshman, A.~Mazumdar, and S.~Bose, ``{Qudits for witnessing
  quantum-gravity-induced entanglement of masses under decoherence},'' {\em
  Phys. Rev. A}, vol.~104, no.~5, p.~052416, 2021.

\bibitem{Schut:2021svd}
M.~Schut, J.~Tilly, R.~J. Marshman, S.~Bose, and A.~Mazumdar, ``{Improving
  resilience of quantum-gravity-induced entanglement of masses to decoherence
  using three superpositions},'' {\em Phys. Rev. A}, vol.~105, no.~3,
  p.~032411, 2022.

\bibitem{Grossardt:2020def}
A.~Gro\ss{}ardt, ``{Acceleration noise constraints on gravity induced
  entanglement},'' {\em Phys. Rev. A}, vol.~102, no.~4, p.~040202, 2020.

\bibitem{Toros:2020krn}
M.~Toro\v{s}, A.~Mazumdar, and S.~Bose, ``{Loss of coherence of matter-wave
  interferometer from fluctuating graviton bath},'' 8 2020.

\bibitem{Wu:2022rdv}
M.-Z. Wu, M.~Toro\v{s}, S.~Bose, and A.~Mazumdar, ``{Quantum gravitational
  sensor for space debris},'' {\em Phys. Rev. D}, vol.~107, no.~10, p.~104053,
  2023.

\bibitem{Gunnink:2022ner}
F.~Gunnink, A.~Mazumdar, M.~Schut, and M.~Toro\v{s}, ``{Gravitational
  decoherence by the apparatus in the quantum-gravity induced entanglement of
  masses},'' 10 2022.

\bibitem{Nguyen:2019huk}
H.~C. Nguyen and F.~Bernards, ``{Entanglement dynamics of two mesoscopic
  objects with gravitational interaction},'' {\em The European Physical Journal
  D}, vol.~74, pp.~1--5, 6 2020.

\bibitem{Rijavec:2020qxd}
S.~Rijavec, M.~Carlesso, A.~Bassi, V.~Vedral, and C.~Marletto, ``{Decoherence
  effects in non-classicality tests of gravity},'' {\em New J. Phys.}, vol.~23,
  no.~4, p.~043040, 2021.

\bibitem{Romero_Isart_2011}
O.~Romero-Isart, ``Quantum superposition of massive objects and collapse
  models,'' {\em Physical Review A}, vol.~84, nov 2011.

\bibitem{Chang_2009}
D.~E. Chang, C.~A. Regal, S.~B. Papp, D.~J. Wilson, J.~Ye, O.~Painter, H.~J.
  Kimble, and P.~Zoller, ``Cavity opto-mechanics using an optically levitated
  nanosphere,'' {\em Proceedings of the National Academy of Sciences},
  vol.~107, pp.~1005--1010, dec 2009.

\bibitem{Sinha:2022snc}
K.~Sinha and P.~W. Milonni, ``{Dipoles in blackbody radiation: momentum
  fluctuations, decoherence, and drag force},'' {\em J. Phys. B}, vol.~55,
  no.~20, p.~204002, 2022.

\bibitem{Deli__2020}
U.~Deli{\'{c}}, M.~Reisenbauer, K.~Dare, D.~Grass, V.~Vuleti{\'{c}}, N.~Kiesel,
  and M.~Aspelmeyer, ``Cooling of a levitated nanoparticle to the motional
  quantum ground state,'' {\em Science}, vol.~367, pp.~892--895, feb 2020.

\bibitem{Tebbenjohanns_2020}
F.~Tebbenjohanns, M.~Frimmer, V.~Jain, D.~Windey, and L.~Novotny, ``Motional
  sideband asymmetry of a nanoparticle optically levitated in free space,''
  {\em Physical Review Letters}, vol.~124, jan 2020.

\bibitem{Whittle:2021mtt}
C.~Whittle {\em et~al.}, ``{Approaching the motional ground state of a 10-kg
  object},'' {\em Science}, vol.~372, no.~6548, pp.~1333--1336, 2021.

\bibitem{Miao:2019pxw}
H.~Miao, D.~Martynov, H.~Yang, and A.~Datta, ``{Quantum correlation of light
  mediated by gravity},'' {\em Phys. Rev. A}, vol.~101, no.~6, p.~063804, 2020.

\bibitem{Miki:2020hvg}
D.~Miki, A.~Matsumura, and K.~Yamamoto, ``{Entanglement and decoherence of
  massive particles due to gravity},'' {\em Phys. Rev. D}, vol.~103, no.~2,
  p.~026017, 2021.

\bibitem{Hsu:2016}
J.-F. Hsu, P.~Ji, C.~W. Lewandowski, and B.~D’Urso, ``Cooling the motion of
  diamond nanocrystals in a magneto-gravitational trap in high vacuum,'' {\em
  Scientific reports}, vol.~6, no.~1, p.~30125, 2016.

\bibitem{WoodPRB22_GM}
B.~D. Wood {\em et~al.}, ``Long spin coherence times of nitrogen vacancy
  centers in milled nanodiamonds,'' {\em Phys. Rev. B}, vol.~105, p.~205401,
  2022.

\bibitem{Doherty_2013}
M.~W. Doherty, N.~B. Manson, P.~Delaney, F.~Jelezko, J.~Wrachtrup, and L.~C.
  Hollenberg, ``The nitrogen-vacancy colour centre in diamond,'' {\em Physics
  Reports}, vol.~528, pp.~1--45, jul 2013.

\bibitem{Afek:2021bua}
G.~Afek, F.~Monteiro, B.~Siegel, J.~Wang, S.~Dickson, J.~Recoaro, M.~Watts, and
  D.~C. Moore, ``{Control and measurement of electric dipole moments in
  levitated optomechanics},'' {\em Phys. Rev. A}, vol.~104, no.~5, p.~053512,
  2021.

\bibitem{Rider:2016xaq}
A.~D. Rider, D.~C. Moore, C.~P. Blakemore, M.~Louis, M.~Lu, and G.~Gratta,
  ``{Search for Screened Interactions Associated with Dark Energy Below the 100
  $\mathrm{\mu m}$ Length Scale},'' {\em Phys. Rev. Lett.}, vol.~117, no.~10,
  p.~101101, 2016.

\bibitem{Barker:2022mdz}
P.~F. Barker, S.~Bose, R.~J. Marshman, and A.~Mazumdar, ``{Entanglement based
  tomography to probe new macroscopic forces},'' {\em Phys. Rev. D}, vol.~106,
  no.~4, p.~L041901, 2022.

\bibitem{Casimir:1947kzi}
H.~B.~G. Casimir and D.~Polder, ``{The Influence of retardation on the
  London-van der Waals forces},'' {\em Phys. Rev.}, vol.~73, pp.~360--372,
  1948.

\bibitem{Casimir:1948dh}
H.~B.~G. Casimir, ``{On the Attraction Between Two Perfectly Conducting
  Plates},'' {\em Indag. Math.}, vol.~10, pp.~261--263, 1948.

\bibitem{griffiths2005introduction}
D.~J. Griffiths, ``Introduction to electrodynamics,'' 2005.

\bibitem{feynman1963feynman}
R.~Feynman, R.~Leighton, and M.~Sands, ``Feynman lectures on physics. vol. 2:
  Mainly electromagnetism and matter, 592 pp,'' 1963.

\bibitem{kim2005static}
H.-Y. Kim, J.~O. Sofo, D.~Velegol, M.~W. Cole, and G.~Mukhopadhyay, ``Static
  polarizabilities of dielectric nanoclusters,'' {\em Physical Review A},
  vol.~72, no.~5, p.~053201, 2005.

\bibitem{pavlou2021effective}
C.~Pavlou, M.~G. Pastore~Carbone, A.~C. Manikas, G.~Trakakis, C.~Koral,
  G.~Papari, A.~Andreone, and C.~Galiotis, ``Effective emi shielding behaviour
  of thin graphene/pmma nanolaminates in the thz range,'' {\em Nature
  Communications}, vol.~12, no.~1, p.~4655, 2021.

\bibitem{Floch2011:EPO}
J.-M.~L. Floch, R.~Bara, J.~G. Hartnett, M.~E. Tobar, D.~Mouneyrac,
  D.~Passerieux, D.~Cros, J.~Krupka, P.~Goy, and S.~Caroopen, ``Electromagnetic
  properties of polycrystalline diamond from 35 k to room temperature and
  microwave to terahertz frequencies,'' {\em Journal of Applied Physics},
  vol.~109, no.~9, p.~094103, 2011.

\bibitem{Ye2005:die}
H.~Ye, H.~Yan, and R.~B. Jackman, ``Dielectric properties of single crystal
  diamond,'' {\em Semiconductor science and technology}, vol.~20, no.~3,
  p.~296, 2005.

\bibitem{Ibaraa1997:wide}
A.~Ibarra, M.~Gonzalez, R.~Vila, and J.~Molla, ``Wide frequency dielectric
  properties of cvd diamond,'' {\em Diamond and Related Materials}, vol.~6,
  no.~5-7, pp.~856--859, 1997.

\bibitem{Ford:1998ex}
L.~H. Ford, ``Casimir force between a dielectric sphere and a wall: A model for
  amplification of vacuum fluctuations,'' {\em Phys. Rev. A}, vol.~58,
  pp.~4279--4286, Dec 1998.

\bibitem{Rivic}
F.~Rivic, A.~Lehr, and R.~Schäfer, ``Scaling of the permanent electric dipole
  moment in isolated silicon clusters with near-spherical shape,'' {\em
  Physical chemistry chemical physics : PCCP}, vol.~25, 05 2023.

\bibitem{Subramaniam2013:one}
C.~Subramaniam, T.~Yamada, K.~Kobashi, A.~Sekiguchi, D.~N. Futaba, M.~Yumura,
  and K.~Hata, ``One hundred fold increase in current carrying capacity in a
  carbon nanotube--copper composite,'' {\em Nature communications}, vol.~4,
  no.~1, p.~2202, 2013.

\bibitem{Schlosshauer:2019ewh}
M.~Schlosshauer, ``{Quantum Decoherence},'' {\em Phys. Rept.}, vol.~831,
  pp.~1--57, 2019.

\bibitem{Toros:2020dbf}
M.~Toro\v{s}, T.~W. van~de Kamp, R.~J. Marshman, M.~S. Kim, A.~Mazumdar, and
  S.~Bose, ``{Relative acceleration noise mitigation for nanocrystal
  matter-wave interferometry: Applications to entangling masses via quantum
  gravity},'' {\em Phys. Rev. Res.}, vol.~3, no.~2, p.~023178, 2021.

\bibitem{Geraci:atomchip}
C.~Montoya, J.~Valencia, A.~A. Geraci, M.~Eardley, J.~Moreland, L.~Hollberg,
  and J.~Kitching, ``Resonant interaction of trapped cold atoms with a magnetic
  cantilever tip,'' {\em Phys. Rev. A}, vol.~91, p.~063835, Jun 2015.

\end{thebibliography}
\bibliographystyle{ieeetr}
\end{document}